\documentclass[12pt]{article}
\usepackage{epsfig}
\usepackage{amssymb}
\usepackage{amsmath}
\usepackage{amsfonts}
\usepackage{graphicx}
\usepackage{mathrsfs}
\DeclareMathAlphabet{\mathscrbf}{OMS}{mdugm}{b}{n}
\usepackage{mathabx}
\usepackage[dvips]{color}
\usepackage{multirow}
\usepackage{calc}
\usepackage{accents}

\newcommand{\vardbtilde}[1]{\tilde{\raisebox{0pt}[0.95\height]{$\tilde{#1}$}}}

\newcommand{\bsigma}{\boldsymbol{\sigma}}

\newcommand{\bnabla}{\boldsymbol{\nabla}}

\newcommand{\R}{\mathbb{R}}
\newcommand{\C}{\mathbb{C}}

\newcommand{\fa}{\mathfrak{a}}

\newcommand{\fp}{\mathfrak{p}}

\newcommand{\fs}{\mathfrak{s}}

\newcommand{\fz}{\mathfrak{z}}

\newcommand{\fB}{\mathfrak{B}}

\newcommand{\fJ}{\mathfrak{J}}
\newcommand{\fK}{\mathfrak{K}}

\newcommand{\fR}{\mathfrak{R}}
\newcommand{\fS}{\mathfrak{S}}
\newcommand{\fT}{\mathfrak{T}}

\newcommand{\fZ}{\mathfrak{Z}}

\newcommand{\bfa}{\mathbf{a}}

\newcommand{\bc}{\mathbf{c}}

\newcommand{\bfe}{\mathbf{e}}

\newcommand{\bg}{\mathbf{g}}

\newcommand{\bk}{\mathbf{k}}
\newcommand{\bj}{\mathbf{j}}

\newcommand{\bp}{{\mathbf{p}}}
\newcommand{\bfr}{\mathbf{r}}
\newcommand{\bt}{\mathbf{t}}

\newcommand{\bA}{\mathbf{A}}
\newcommand{\bcA}{\boldsymbol{\cA}}
\newcommand{\bB}{\mathbf{B}}
\newcommand{\bC}{\mathbf{C}}
\newcommand{\bD}{\mathbf{D}}

\newcommand{\bcE}{{\boldsymbol{\cE}}}
\newcommand{\bF}{\mathbf{F}}

\newcommand{\bH}{\mathbf{H}}
\newcommand{\bI}{\mathbf{I}}
\newcommand{\bJ}{\mathbf{J}}
\newcommand{\bK}{\mathbf{K}}
\newcommand{\bL}{\mathbf{L}}
\newcommand{\bcL}{\boldsymbol{\cL}}
\newcommand{\bM}{\mathbf{M}}

\newcommand{\bS}{\mathbf{S}}
\newcommand{\bT}{\mathbf{T}}

\newcommand{\bV}{\mathbf{V}}

\newcommand{\cA}{{\mathcal{A}}}
\newcommand{\cB}{\mathcal{B}}
\newcommand{\cC}{\mathcal{C}}

\newcommand{\cH}{\mathcal{H}}
\newcommand{\cE}{\mathcal{E}}
\newcommand{\cF}{\mathcal{F}}

\newcommand{\cJ}{\mathcal{J}}
\newcommand{\bcJ}{{\boldsymbol{\cJ}}}
\newcommand{\cK}{\mathcal{K}}
\newcommand{\cL}{\mathcal{L}}
\newcommand{\cM}{\mathcal{M}}

\newcommand{\cR}{\mathcal{R}}
\newcommand{\cS}{\mathcal{S}}

\newcommand{\cU}{\mathcal{U}}

\newcommand{\cZ}{\mathcal{Z}}
\newcommand{\be}{\begin{equation}}
\newcommand{\ee}{\end{equation}}
\newcommand{\bea}{\begin{eqnarray}}
\newcommand{\eea}{\end{eqnarray}}
\newcommand{\nn}{\nonumber}
\newcommand{\kt}{\rangle}
\newcommand{\br}{\langle}

\newcommand{\ed}{\end{document}}

\newcommand{\bi}{\begin{itemize}}
\newcommand{\ei}{\end{itemize}}

\newcommand{\bce}{\begin{center}}
\newcommand{\ece}{\end{center}}

\newcommand{\sD}{\mathscr{D}}

\newcommand{\sF}{\mathscr{F}}
\newcommand{\sG}{\mathscr{G}}
\newcommand{\sH}{\mathscr{H}}

\newcommand{\sR}{\mathscr{R}}

\newcommand{\bsS}{\mathscrbf{S}}

\newcommand{\bsZ}{\mathscrbf{Z}}

\newcommand{\RE}{{\rm Re}}
\newcommand{\IM}{{\rm Im}}

\newcommand{\bPsi}{{\boldsymbol{\Psi}}}
\newcommand{\bPhi}{{\boldsymbol{\Phi}}}
\newcommand{\bPi}{{\boldsymbol{\Pi}}}
\newcommand{\bcB}{{\boldsymbol{\cB}}}
\newcommand{\bfB}{{\boldsymbol{\fB}}}
\newcommand{\bcC}{{\boldsymbol{\cC}}}

\newcommand{\bcK}{{\boldsymbol{\cK}}}
\newcommand{\bcM}{{\boldsymbol{\cM}}}

\newcommand{\bfJ}{{\boldsymbol{\fJ}}}
\newcommand{\bcH}{{\boldsymbol{\cH}}}
\newcommand{\bcU}{{\boldsymbol{\cU}}}
\newcommand{\bfZ}{{\boldsymbol{\fZ}}}
\newcommand{\bvarepsilon}{{{\mbox{\large$\boldsymbol{\varepsilon}$}}}}
\newcommand{\bmu}{{{\mbox{\large$\boldsymbol{\mu}$}}}}

\newcommand{\bigvarepsilon}{\mbox{\large$\varepsilon$}}
\newcommand{\bigmu}{\mbox{\large$\mu$}}

\newcommand{\bzero}{{\boldsymbol{0}}}
\newcommand{\p}{\boldsymbol{\partial}}

\newcommand{\for}{{\mbox{\rm for}}}

\newcommand{\bXi}{{\boldsymbol{\Xi}}}

\newcommand{\Lpi}{{\widehat{\mbox{\Large$\pi$}}_{\!k}}}

\newcommand{\msF}{{\mathring\sF}}

\newcommand{\bfp}{{\boldsymbol{\fp}}}

\begin{document}


\title{Introducing a general method for solving electromagnetic radiation problem in an arbitrary linear medium}

\author{Farhang Loran\thanks{E-mail address: loran@iut.ac.ir}
~and Ali~Mostafazadeh\thanks{E-mail address:
amostafazadeh@ku.edu.tr} $^{,\,\ddagger}$\\[6pt]
$^*$Department of Physics, Isfahan University of Technology, \\ Isfahan 84156-83111, Iran\\[6pt]
$^\dagger$Departments of Mathematics and Physics, Ko\c{c} University,\\  34450 Sar{\i}yer,
Istanbul, T\"urkiye\\[6pt]
$^\ddagger$T\"{U}B$\dot{\rm I}$TAK Research Institute for Fundamental Sciences,\\ Gebze, Kocaeli 41470, T\"urkiye}

\date{ }
\maketitle

\begin{abstract}

Numerical transfer matrices have been widely used in the study of wave propagation and scattering. These may be viewed as descretizations of a recently introduced fundamental notion of transfer matrix which admits a representation in terms of the evolution operator for an effective non-unitary quantum system.
We use the fundamental transfer matrix to develop a general method for the solution of the problem of radiation of an oscillating source in an arbitrary, possibly non-homogenous, anisotropic, and active or lossy linear medium. This allows us to obtain an analytic solution of this problem for an oscillating source located in the vicinity of a planar collection of possibly anisotropic and active/lossy point scatterers such as those modeling a two-dimensional photonic crystal.


\vspace{2mm}



\end{abstract}

\section{Introduction}
\label{S1}

Electromagnetic radiation of an oscillating source is a physical phenomenon of great importance. By definition, a system of charges and currents radiates if it generates waves reaching spatial infinities, i.e., they are detectable by detectors located far away from the source \cite{Jackson}. This is in contrast to the basic setup for a scattering problem where not only the detectors but the source of the wave reside at spatial infinities \cite{Newton}. A more realistic situation is when the waves generated by a source interact with nearby scatterers before reaching the detectors. The purpose of this article is to develop a general method of dealing with this problem which is particularly effective for the description of the effects of the point scatterers on the emitted radiation.

The term ``point scatterer'' refers to an interaction with a negligibly small  (zero) range \cite{Albaverio}. The best-known examples are the interactions modeled by delta-function potentials. These have been extensively studied since the 1930's \cite{KP,Fermi,Foldy-1945,Lieb-1963a,Lieb-1963b,Antonie-1994,cervero,batchelor,josa-2020}. In one dimension, they provide useful exactly solvable toy models with interesting physical applications \cite{flugge,tjp-2000}. In two and higher dimensions, their standard treatment 
leads to divergent terms whose removal requires a coupling-constant renormalization \cite{thorn,jackiw,mead,manuel,Adhikari1,Adhikari2,Mitra,ap-2019}.
The same problem arises in the study of the scattering of electromagnetic waves by delta-function permittivity profiles and leads to more serious complications even when they are isotropic \cite{VCL,calla-2014}. 

Recently, we have developed an alternative approach to the scattering of scalar and electromagnetic waves which avoids the singularities of the standard treatment of point scatterers provided that they lie along a line in two dimensions and on a plane in three dimensions \cite{pra-2021,ap-2022b,pra-2023}. This approach is based on a fundamental notion of transfer matrix which unlike the transfer matrices employed in the earlier publications \cite{pendry-1984,pendry-1990a,pendry-1996} allows for performing analytic calculations. This has so far led to the discovery of exact broadband unidirectional invisibility in two dimensions \cite{ol-2017} and the construction of potentials for which the first Born approximation is exact \cite{pra-2019}. See also \cite{pra-2017}. These developments together with the remarkable effectiveness of the fundamental transfer matrix in dealing with point scatterers provide the basic motivation for exploring its utility in dealing with radiation problems.

The outline of this article is as follows. In Sec.~\ref{S2}, we give the definition of the fundamental transfer matrix for a general (possibly non-homogenous, anisotropic, active, or lossy) stationary linear medium that contains an oscillating localized distribution of charges and currents. In Sec.~\ref{S3}, we discuss the application of the fundamental transfer matrix in addressing the radiation problem for this setup. In Sec.~\ref{S4}, we address the problem of radiation of an oscillating source in the presence of a finite planar array of non-magnetic point interactions. In Sec.~\ref{S5} we confine our attention to the case where the source is a perfect dipole. Here we also explore in some detail the special case where the radiation of the dipole is affected by the presence of a single point scatterer. In Sec.~\ref{S6} we present our concluding remarks.

\section{Fundamental transfer matrix for electromagnetic waves}
\label{S2}

Consider a stationary linear medium that contains an oscillating localized distribution of charges and currents  (the source). Let $\bvarepsilon$ and $\bmu$ denote the  permittivity and permeability tensors of the medium, $\varepsilon_0$ and $\mu_0$ be the permittivity and permeability of vacuum, and $\omega$ be the angular frequency of the source. Then $\bvarepsilon$ and $\bmu$ are $3\times 3$ matrix-valued functions of space, and we can respectively express the free charge and current densities of the source, and the electric and magnetic fields of the generated wave as
	\begin{align}
	&\rho(\bfr,t)=\sqrt{\varepsilon_0}\,e^{-i\omega t}\varrho(\bfr),
	&&\boldsymbol{\mathsf{J}}(\bfr,t)=\frac{e^{-i\omega t}\bcJ(\bfr)}{\sqrt{\mu_0}},
	\label{rho-sJ}\\
	&\boldsymbol{\mathsf{E}}(\bfr,t)=\frac{e^{-i\omega t}\bcE(\bfr)}{\sqrt{\varepsilon_0}},
	&&\boldsymbol{\mathsf{H}}(\bfr,t)=\frac{e^{-i\omega t}\bcH(\bfr)}{\sqrt{\mu_0}},
	\label{sE-sH}
	\end{align} 
where $\bfr$ stands for the position vector, $\varrho$ is a scalar function, and $\bcJ$, $\bcE$, and $\bcH$ are vector-valued functions. In terms of these, Maxwell's equations \cite{Jackson} take the form
	\begin{align}
	&\bnabla\cdot(\hat\bvarepsilon\,\bcE)=\varrho,
	&&\bnabla\cdot(\hat\bmu\,\bcH)=0,
	\label{mx1}\\
	&\bnabla\times\bcE=ik\hat\bmu\,\bcH,
	&&\bnabla\times\bcH=-ik\hat\bvarepsilon\,\bcE+\bcJ,
	\label{mx2}
	\end{align}
where $\hat\bvarepsilon$$(\bfr):=\varepsilon_0^{-1}\bvarepsilon(\bfr)$ and $\hat\bmu$$(\bfr):=\mu_0^{-1}\bmu(\bfr)$ are respectively the relative permittivity and permeability tensors, $k:=\omega/c$ is the wavenumber, and $c:=(\varepsilon_0\mu_0)^{-1/2}$ is the speed of light in vacuum.\footnote{According to the continuity equation for the local charge conservation, which follows from the first equation in (\ref{mx1}) and the second equation in (\ref{mx2}), we have $\varrho=-ik^{-1}\bnabla\cdot\bcJ$. Therefore, $\bcJ$ characterizes the source.}

\begin{figure}
        \begin{center}
        \includegraphics[scale=.25]{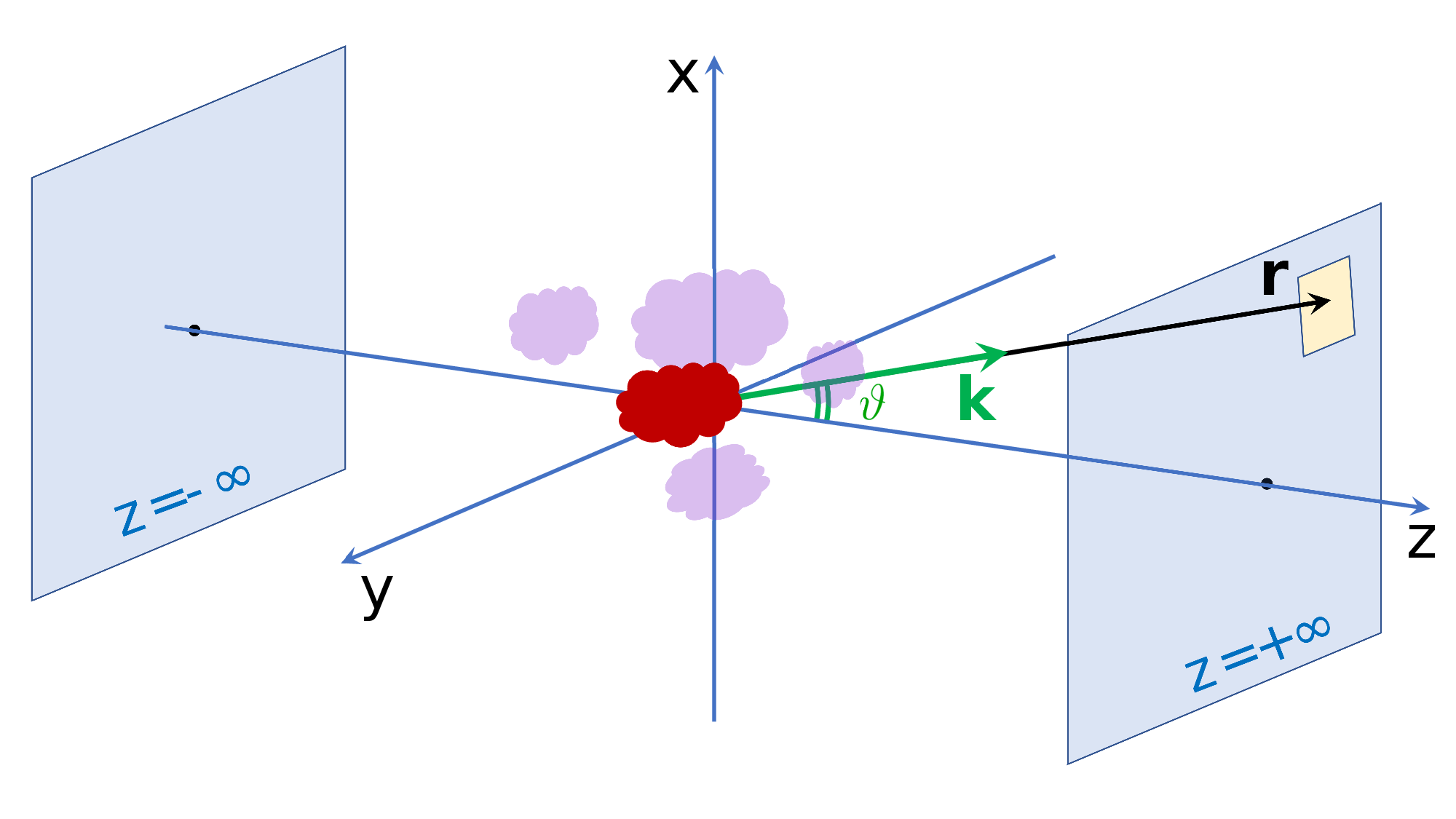}
        \caption{Schematic view of the setup for the radiation of an oscillating source (painted in red) in a linear medium containing scatterers (non-homogeneous, anisotropic, active or lossy regions marked in purple). The detectors are placed on the planes $z=\pm\infty$. $\bfr$ is the position of a detector's screen (painted in yellow) that lies on the plane $z=+\infty$. $\bk$ is the wave vector for the detected wave. $\vartheta$ is the polar angle of the spherical coordinates.}
        \label{fig1}
        \end{center}
        \end{figure}

Ref.~\cite{pra-2023} defines the fundamental transfer matrix for a general linear medium in the absence of free charges and currents. We wish to extend this definition to linear media containing an oscillating source. To do this, we choose our coordinate system in such a way that the detectors measuring the radiation lie on the planes defined by $z=\pm\infty$, as depicted in Fig.~\ref{fig1}.
We also suppose that the last diagonal entry of $\hat\bvarepsilon$ and $\hat\bmu$ do not vanish; $\hat\varepsilon_{33}\neq 0\neq\hat\mu_{33}$. This is a technical condition which we can satisfy by a proper choice of our coordinate system for nonexotic media.

In the following, we denote the zero and identity matrices of all sizes by $\bzero$ and $\bI$, respectively, and use $\cE_u$, $\cH_u$, and $\cJ_u$, with $u\in\{x,y,z\}$, to denote the components of $\bcE$, $\bcH$,  and $\bcJ$, i.e., 
	\begin{align*}
	&\bcE=\cE_x\bfe_x+\cE_y\bfe_y+\cE_z\bfe_z, 
	&&\bcH=\cH_x\bfe_x+\cH_y\bfe_y+\cH_z\bfe_z,
	&&\bcJ=\cJ_x\bfe_x+\cJ_y\bfe_y+\cJ_z\bfe_z,
	\end{align*} 
where $\bfe_u$ is the unit vector pointing along the $u$-axis. We also introduce the following quantities.\footnote{In Ref.~\cite{pra-2023} we use $\vec J_\cE, \bJ_\cE,\vec J_\cH$, and $\bJ_\cH$ for what we call $\vec K_\cE, \bK_\cE,\vec K_\cH$, and $\bK_\cH$. This change of notation has been made to avoid giving the impression that these quantities are related to the current density $\boldsymbol{\mathsf{J}}$.}
	\begin{align}
	&\vec\cE :=\left[\begin{array}{c}
	\cE_x \\ \cE_y \end{array}\right], 
	\quad\quad\quad
	\vec\cH :=\left[\begin{array}{c}
	\cH_x \\ \cH_y \end{array}\right],
	\quad\quad\quad
	\vec\cJ:=\left[\begin{array}{c}
	\cJ_x \\ \cJ_y \end{array}\right],
	\quad\quad\quad
	\vec\partial:=\left[\begin{array}{c}\partial_x \\ \partial_y \end{array}\right],\\
	&\vec K_\cE:=\Bigg[\begin{array}{c}
	-\hat\varepsilon_{23}\\
	\hat\varepsilon_{13}\end{array}\Bigg],
	\quad\quad\quad\quad
	\bK_\cE:=\Bigg[\begin{array}{cc}
	-\hat\varepsilon_{21} & -\hat\varepsilon_{22}\\
	\hat\varepsilon_{11} & \hat\varepsilon_{12}
	\end{array}\Bigg]=\Bigg[\begin{array}{c}
	-\vec\bigvarepsilon_2^{\,T}\\
	\vec\bigvarepsilon_1^{\,T}\end{array}\Bigg],
	\label{KEs=}\\
	&\vec K_\cH:=\Bigg[\begin{array}{c}
	-\hat\mu_{23}\\
	\hat\mu_{13}\end{array}\Bigg],
	\quad\quad\quad\quad
	\bK_\cH:=
	\Bigg[\begin{array}{cc}
	-\hat\mu_{21} & -\hat\mu_{22}\\
	\hat\mu_{11} & \hat\mu_{12}
	\end{array}\Bigg]=\Bigg[\begin{array}{c}
	-\vec\bigmu_2^{\,T}\\
	\vec\bigmu_1^{\,T}\end{array}\Bigg],
	\label{KHs=}
	\end{align}
where the superscript $T$ stands for the transpose of the corresponding matrix, and
	\begin{align}
	&\vec\bigvarepsilon_\ell :=\left[\begin{array}{c}
	\hat\varepsilon_{\ell 1} \\
	\hat\varepsilon_{\ell 2} \end{array}\right],
	&&
	\vec\bigmu_\ell :=\left[\begin{array}{c}
	\hat\mu_{\ell 1} \\
	\hat\mu_{\ell 2} \end{array}\right],
	&&\ell\in\{1,2,3\}.\nn
	\end{align}
	
We begin our analysis by using (\ref{mx2}) to express $\cE_z$ and $\cH_z$ in the form
	\begin{align}
	\cE_z&=\hat\varepsilon_{33}^{-1}\left[
	-\hat\varepsilon_{31}\cE_x-\hat\varepsilon_{32}\cE_y+
	ik^{-1}\left(\partial_x\cH_y-\partial_y\cH_x-\cJ_z\right)\right]\nn\\
	&=-\hat\varepsilon_{33}^{-1}\Big(\vec\bigvarepsilon_3^{\;T}\vec\cE
	+k^{-1}\vec\partial^{\;T}\bsigma_2\vec\cH+ik^{-1}\cJ_z\Big),
	\label{Ez=}\\[3pt]
	\cH_z&=\hat\mu_{33}^{-1}\left[
	-ik^{-1}\left(\partial_x\cE_y-\partial_y\cE_x\right)
	-\hat\mu_{31}\cH_x-\hat\mu_{32}\cH_y\right]\nn\\
	&=\hat\mu_{33}^{-1}\Big(
	k^{-1}\vec\partial^{\;T}\bsigma_2\,\vec\cE
	-\vec\bigmu_3^{\;T}\vec\cH\Big).
	\label{Hz=}
	\end{align}

With the help of (\ref{Ez=}) and (\ref{Hz=}), we  can reduce (\ref{mx2}) to a system of first-order differential equations for $\cE_x,\cE_y,\cH_x$ and $\cH_y$. This is equivalent to the non-homogenous time-dependent Schr\"odinger equation,
	\be
	i\partial_z\bPhi=\widehat\bH\,\bPhi+\boldsymbol{\bJ},
	\label{TD-sch-eq}
	\ee
for the $4$-component field \cite{pra-2023,berreman-1972},
	\be
	\bPhi :=\left[\begin{array}{c}
	\cE_x \\ \cE_y \\ 
	\cH_x  \\ \cH_y 	
	\end{array}\right]=\left[\begin{array}{c}
	\vec\cE \\ \vec\cH \end{array}\right],
	\label{4-com-field}
	\ee
where $z$ plays the role of `time,' 
	\begin{align}
	&\widehat\bH:=\left[\begin{array}{cc}
	\widehat\bH_{11} & \widehat\bH_{12} \\[6pt]
	\widehat\bH_{21}  & \widehat\bH_{22}\end{array}\right],
	&&\bJ:=\left[\begin{array}{c}
	k^{-1}\vec\p\left( \hat\varepsilon_{33}^{\ -1}\cJ_z\right)\\[3pt] 
	i\hat\varepsilon_{33}^{\ -1}\cJ_z\vec K_\cE-\bsigma_2\vec\cJ\end{array}\right],
	\label{H-J-def}\\
	&{\widehat\bH}_{11}:=
	-i{\vec\partial}\;
	\frac{\vec\bigvarepsilon_3^{\,T}}{\hat\varepsilon_{33}}+
	\frac{1}{\hat\mu_{33}}\vec K_\cH\vec\partial^{\,T}\bsigma_2,
    	&&{\widehat\bH}_{12}:=
	-\frac{i}{k}{\vec\partial}\;\frac{1}{\hat\varepsilon_{33}}
	\vec\partial^{\,T}\bsigma_2 +k(\bK_\cH-\breve\bK_\cH),
	\label{H1}\\
    	&{\widehat\bH}_{21}:=
	\frac{i}{k}{\vec\partial}\;\frac{1}{\hat\mu_{33}}
	\vec\partial^{\,T}\bsigma_2+ k(\breve\bK_\cE-\bK_\cE),
    	&&{\widehat\bH}_{22}:=
	-i{\vec\partial}\;\frac{\vec\bigmu_3^{\,T}}{\hat\mu_{33}}+
	\frac{1}{\hat\varepsilon_{33}}\vec K_\cE\vec\partial^{\,T}\bsigma_2,
	\label{H2}\\
    	&\breve\bK_\cE:=\frac{1}{\hat\varepsilon_{33}}\vec K_\cE\vec\bigvarepsilon_3^{\,T},
	&&\breve\bK_\cH:=\frac{1}{\hat\mu_{33}}\vec K_\cH\vec\bigmu_3^{\,T},
	\label{tKs=}
	\end{align}
and ${\vec\partial}$ and ${\vec\partial}^T$ act on all the terms appearing to their right.\footnote{For example, for every test function $f$,  ${\vec\partial}\,
	\frac{\vec\bigvarepsilon_3^{\,T}}{\hat\varepsilon_{33}}f$ stands for ${\vec\partial}\big(\frac{\vec\bigvarepsilon_3^{\,T}}{\hat\varepsilon_{33}}f\big)$.}
According to (\ref{H-J-def}) -- (\ref{tKs=}), $\widehat\bH$ is a `time-dependent' $4\times 4$ matrix Hamiltonian with operator entries which represents the interaction of the electromagnetic waves with the medium, $\widehat\bH_{ij}$ are the $2\times 2$ blocks of $\widehat\bH$, and $\bJ$ is a 4-component function that contains the information about the source. 

Because $z$ plays a different role than $x$ and $y$, we denote the projection of $\bfr$ onto the $x$-$y$ plane by $\vec r$, i.e., set $\vec r:=(x,y)$, and write $\bPhi(\bfr)$ as $\bPhi(\vec r,z)$. This allows us to view $\bPhi(\cdot,z)$ as a function that maps $\R^2$ to $\C^{4\times 1}$, where $\C^{m\times n}$ denotes the set of $m\times n$ complex matrices. The Hamiltonian operator $\widehat\bH$ is actually a $z$-dependent linear operator acting in the space of such functions. Eq.~(\ref{TD-sch-eq}) determines a dynamics in this space. 

The $4$-component function $\bPhi(\cdot,z)$ plays the role of a position-space wave function in quantum mechanics. In the following, we employ the corresponding momentum-space wave function $\tilde\bPhi(\cdot,z)$ which is related to $\bPhi(\cdot,z)$ by the two-dimensional Fourier transformation $\cF$; 
	\[\bPhi(\cdot,z)\stackrel{\cF}{\longrightarrow}\cF\bPhi(\cdot,z):=\tilde\bPhi(\cdot,z),\] 
where for all $\vec p:=(p_x,p_y)\in\R^2$,
	\be
	\tilde\bPhi(\vec p,z):=\int_{\R^2}d^2 r\:e^{-i\vec p\cdot\vec r}\bPhi(\vec r,z)=\int_{-\infty}^\infty\!\! dx\int_{-\infty}^\infty\!\!dy\:e^{-i(xp_x+yp_y)}\bPhi(x,y,z),
	\label{t-bPhi=}
	\ee
and a dot stands for the dot product, i.e., $\vec p\cdot\vec r:=xp_x+yp_y$. 
Applying $\cF$ to both sides of (\ref{TD-sch-eq}) and evaluating the resulting equation at $(\vec p,z)$, we find
	\be
	i\partial_z\tilde\bPhi(\vec p,z)=\widehat{\tilde\bH}\,\tilde\bPhi(\vec p,z)+\tilde{\bJ}(\vec p,z),
	\label{TD-sch-eq-FT}
	\ee
where $\widehat{\tilde\bH}:=\cF\,\widehat{\bH}\,\cF^{-1}$, $\cF^{-1}$ stands for the inverse Fourier transformation in two dimensions\footnote{Given a test function $\bF:\R^2\to\C^{m\times n}$, $\big(\cF^{-1}\bF\big)(\vec r):=\frac{1}{4\pi^2}\int_{\R^2}d^2p\: e^{i\vec p\cdot\vec r}\bF(\vec p)$.}, and $\tilde\bJ(\cdot,z):=\cF\bJ(\cdot,z)$.\footnote{We can express the $\widehat{\tilde\bH}\tilde\bPhi$ appearing in (\ref{TD-sch-eq-FT}) as $\int_{\R^2}d^2q\: \bcK(\vec p,\vec q)\tilde\bPhi(\vec q,z)$, where $\bcK(\vec p,\vec q):=\frac{1}{4\pi^2}\int_{\R^2}d^2r e^{-i\vec p\cdot\vec r}\widehat\bH\, e^{i\vec q\cdot\vec r}$. Because $\widehat\bH$ is a differential operator whose coefficients are functions of both $\vec r$ and $z$, the integral kernel $\bcK(\cdot,\cdot)$ depends on $z$.} If we denote the space of $d$-component complex-valued functions of $\vec p$ by $\sF^d$, so that $\tilde\bPhi(\cdot,z)\in\sF^4$, we can view (\ref{TD-sch-eq-FT}) as a dynamical equation in $\sF^4$.
	
Consider an electromagnetic wave propagating in vacuum, so that $\hat\bvarepsilon=\hat\bmu=\bI$ and $\bJ=\bzero$, and let $\tilde\bPhi_0$ denote the corresponding momentum-space $4$-component wave function. Then (\ref{TD-sch-eq-FT}) becomes
	\be
	i\partial_z\tilde\bPhi_0(\vec p,z)={\tilde\bH}_0(\vec p)\tilde\bPhi_0(\vec p,z),
	\label{TD-sch-eq-FT-zero}
	\ee
where 
	\begin{align}
	&\tilde{\bH}_0(\vec p):=\left[\begin{array}{cc}
    	\bzero & \tilde{\bL}_0(\vec p)\\
    	- \tilde{\bL}_0(\vec p) & \bzero\end{array}\right],
    	&&\tilde{\bL}_0(\vec p):=\frac{1}{k}\left[\begin{array}{cc}
    	-p_xp_y & p_x^2-k^2\\
   	-p_y^2+k^2 & p_xp_y\end{array}\right].
    	\label{t-H-zero}
    	\end{align}  
Because $\widehat{\tilde{\bH}}_0(\vec p)$ does not depend on $z$, we can write the general solution of (\ref{TD-sch-eq-FT-zero}) in the form,
	\be
	\tilde\bPhi_0(\vec p,z)=e^{-iz\tilde\bH_0(\vec p)}\bcC(\vec p),
	\label{gen-sol=}
	\ee
where $\bcC\in\sF^4$ is arbitrary. To obtain a more explicit expression for the right-hand side of (\ref{gen-sol=}), we first note that 
	\be
	\tilde\bL_0(\vec p)^2=-\varpi(\vec p)^2\bI,
	\label{L2-id}
	\ee
which in turn implies
	\be
	\tilde\bH_0(\vec p)^2=(k^2-\vec p^{\:2})\bI.
	\label{H2=zero}
	\ee 
For $|\vec p|=k$, this becomes $\tilde\bH_0(\vec p)^2=\bzero$, and we have
	\be
	e^{-iz\tilde\bH_0(\vec p)}=\bI-iz\tilde\bH_0(\vec p).
	\label{exp=1}
	\ee
For $|\vec p|\neq k$, we can expand $e^{-iz\tilde\bH_0(\vec p)}$ in powers of $z\tilde\bH_0(\vec p)$ and use (\ref{H2=zero}) to show that 
	\be
	e^{-iz\tilde\bH_0(\vec p)}=\cos[z\varpi(\vec p)]\,\bI-
	\frac{i\sin[z\varpi(\vec p)]}{\varpi(\vec p)}\,\tilde\bH_0(\vec p),
	\label{exp=2}
	\ee
where $\varpi:\R^2\to\C$ is the function defined by
	\be
	\varpi(\vec p):=\left\{\begin{array}{ccc}
	\sqrt{k^2-\vec p^{\:2}} & \for & |\vec p|<k,\\[2pt]
	i\sqrt{\vec p^{\:2}-k^2} & \for & |\vec p|\geq k.
	\end{array}\right.
	\label{varpi-def}
	\ee
	
In the limit $|\vec p|\to k$, $\varpi(\vec p)$ tends to $0$, and (\ref{exp=2}) reproduces (\ref{exp=1}). Therefore we can determine the value of $\tilde\bPhi_0(\vec p,z)$ for $|\vec p|=k$ by evaluating its $|\vec p|\to k$ limit. This observation allows us to confine our attention to $|\vec p|\neq k$ where $\tilde\bH_0(\vec p)$ is a diagonalizable matrix.  In other words, without loss of generality we can restrict the domain of the definition of $\tilde\bPhi_0(\cdot,z)$ to $\R^2\setminus S^1_k$ where $S^1_k:=\{\vec p\in\R^2\,|\,|\vec p|=k\}$. Denoting the set of $d$-component functions defined on $\R^2\setminus S^1_k$ by $\mathring\sF^d_k$, we identify $\tilde\bPhi_0(\cdot,z)$ with an element of $\mathring\sF^4_k$ and write (\ref{gen-sol=}) in the form
	\be
	\tilde\bPhi_0(\cdot,z)=e^{-iz\widehat{\tilde\bH}_0}\bcC,
	\label{gen-sol=1}
	\ee
where $\widehat{\tilde\bH}_0:\mathring\sF^4_k\to\mathring\sF^4_k$ is the linear operator defined by $\widehat{\tilde\bH}_0\bF(\vec p):={\tilde\bH}_0(\vec p)\bF(\vec p)$, and $\bcC\in\msF_k^4$ is arbitrary.
	
As we mentioned above, for $|\vec p|\neq k$, $\tilde\bH_0(\vec p)$ is a diagonalizable matrix. Equations~(\ref{H2=zero}) and (\ref{varpi-def}) suggest that its spectrum consists of $\pm\varpi(\vec p)$.  We can construct the following projection matrices onto its eigenspaces \cite{pra-2023}.
	\be
	\bPi_j(\vec p):=\frac{1}{2}\left[\bI
    	+\frac{(-1)^j}{\varpi(\vec p)}\,\tilde{\bH}_0(\vec p)\right]
	=\frac{1}{2}\left[
	\begin{array}{cc}
	\bI&\frac{(-1)^j\tilde\bL_0(\vec p)}{\varpi(\vec p)}\\
	\frac{(-1)^{j+1}\tilde\bL_0(\vec p)}{\varpi(\vec p)}&\bI
	\end{array}\right],
    	\label{proj}
    	\ee
where $j\in\{1,2\}$. It is easy to check that 
	\bea
	&&\tilde\bH_0\bPi_j(\vec p)=(-1)^j\varpi(\vec p)\bPi_j(\vec p),
	\label{proj-H0}\\
	&&\bPi_i(\vec p)\bPi_j(\vec p)=\delta_{ij}\bPi_j(\vec p),
	\label{ortho-project}\\
	&&\bPi_1(\vec p)+\bPi_2(\vec p)=\bI,
	\label{complete-proj}
	\eea
where $\delta_{ij}$ stands for the Kronecker delta symbol. These in turn imply
	\bea
	e^{-iz\tilde\bH_0(\vec p)}&=&e^{iz\varpi(\vec p)}\bPi_1(\vec p)+
	e^{-iz\varpi(\vec p)}\bPi_2(\vec p).
	\label{exp=proj}
	\eea
	
Another consequence of (\ref{proj-H0}) and (\ref{complete-proj}) is that for every $\bcC\in\msF^4$, the elements $\bcA$ and $\bcB$ of $\msF^4$ that are given by	
	\begin{align}	
	&\bcA(\vec p):=\bPi_1(\vec p)\bcC(\vec p),
	&&\bcB(\vec p):=\bPi_2(\vec p)\bcC(\vec p),
	\label{cA-cB-def}
	\end{align}
satisfy
	\begin{align}
	&\bcA+\bcB=\bcC,
	\label{AB=C}\\
	&\tilde\bH_0(\vec p)\bcA(\vec p)=-\varpi(\vec p)\bcA(\vec p),
	\label{eg-va1}\\
	&\tilde\bH_0(\vec p)\bcB(\vec p)=\varpi(\vec p)\bcB(\vec p).
	\label{eg-va2}
	\end{align}
Therefore, $\bcA(\vec p)$ and $\bcB(\vec p)$ are respectively eigenvectors of $\tilde\bH_0(\vec p)$ with eigenvalues $-\varpi(\vec p)$ and $\varpi(\vec p)$. 
Furthermore, substituting (\ref{exp=proj}) in (\ref{gen-sol=}) and making use of (\ref{eg-va1}) and (\ref{eg-va2}), we arrive at the following expression for the general solution of (\ref{TD-sch-eq-FT-zero}) for $|\vec p|\neq k$.
	\be
	\tilde\bPhi_0(\vec p,z)=e^{iz\varpi(\vec p)}\bcA(\vec p)+
	e^{-iz\varpi(\vec p)}\bcB(\vec p).
	\label{gen-sol=2}
	\ee
		
Next, we let $\widehat\varpi,\widehat\bPi_j,\Lpi:\msF^4\to\msF^4$ be the operators defined by
	\begin{align}
	&(\widehat\varpi\,\bF)(\vec p):=\varpi(\vec p)\bF(\vec p),
	\label{op-varpi}\\[3pt]
	&(\widehat\bPi_j\,\bF)(\vec p):=\bPi_j(\vec p)\bF(\vec p),
	\label{op-Pi}\\
	&\big(\Lpi\bF\big)(\vec p):=\left\{\begin{array}{ccc}
	\bF(\vec p)&\for&|\vec p|<k,\\
	\bzero&\for&|\vec p|>k,\end{array}\right.
	\label{project}
	\end{align}
where $\vec p\in\R^2\setminus S^1_k$, $j\in\{1,2\}$, and $\bF\in\msF^4$. We can use (\ref{cA-cB-def}) and (\ref{gen-sol=2}) -- (\ref{project}) to establish the following identities.
	\begin{align}
	&[\widehat\varpi,\widehat\bPi_j]=[\widehat\varpi,\Lpi]=
	[\widehat\bPi_j,\Lpi]=
	\widehat\bzero,
	\label{commuting}\\	
	&\bcA=\widehat\bPi_1\bcC,
	\quad\quad\quad\quad\quad
	\bcB=\widehat\bPi_2\bcC,
	\label{cA-cB-def2}\\
	&\tilde\bPhi_0(\cdot,z)=e^{iz\widehat\varpi}\bcA+
	e^{-iz\widehat\varpi}\bcB.
	\label{gen-sol=3}
	\end{align}
	
Suppose that there are $a_\pm\in\R$ with $a_-<a_+$ such that the space outside the region bounded by the planes $z=a_\pm$ is empty. Then, $\hat\bvarepsilon(\vec r,z)-\bI=\hat\bmu(\vec r,z)-\bI=\bJ(\vec r,z)=\bzero$ for $z\notin (a_-,a_+)$, and for every solution of (\ref{TD-sch-eq-FT}) there are $\bcC_\pm\in\msF^4$ such that
	\be
	\tilde\bPhi(\vec p,z)=\left\{\begin{array}{ccc}
	e^{iz\varpi(\vec p)}\bcA_-(\vec p)+
	e^{-iz\varpi(\vec p)}\bcB_-(\vec p)&\for&z<a_-,\\
	e^{iz\varpi(\vec p)}\bcA_+(\vec p)+
	e^{-iz\varpi(\vec p)}\bcB_+(\vec p)&\for&z>a_+,
	\end{array}\right.
	\label{Phi-t-asymp-pm}
	\ee
where	
	\begin{align}	
	&\bcA_\pm:=\widehat\bPi_1\bcC_\pm,
	&&\bcB_\pm:=\widehat\bPi_2\bcC_\pm,
	\label{cA-cB-pm}
	\end{align}
and for $\vec p=\vec p_0\in S^1_k$, the right-hand side of (\ref{Phi-t-asymp-pm}) is to be replaced by its $\vec p\to\vec p_0$ limit. If $\bPhi(\vec r,z)$ is a bounded function of $z$, the same applies to $\tilde\bPhi(\vec p,z)$. This condition together with Eqs.~(\ref{varpi-def}) and (\ref{Phi-t-asymp-pm}), and the facts that $\tilde\bPhi(\vec p,z)$ is a uniformly continuous function of $\vec p$,  
and $e^{\mp iz\varpi(\vec p)}$ blows up for $|\vec p|>k$ as $z\to\pm\infty$ imply that
	\begin{align}
	\bcA_-(\vec p)=\bcB_+(\vec p)=0~~~\for~~~|\vec p|>k.
	\label{asym-condi-1}
	\end{align}

Let us introduce 
	\begin{align}
	&\bC_\pm:=\Lpi\bcC_\pm,
	&&\bA_\pm:=\Lpi\bcA_\pm=\widehat\bPi_1\bC_\pm,
	&&\bB_\pm:=\Lpi\bcB_\pm=\widehat\bPi_2\bC_\pm.
	\label{bABC-def}
	\end{align}
Then, by virtue of (\ref{ortho-project}), (\ref{op-Pi}), (\ref{project}), and (\ref{bABC-def}),
	\begin{align}
	&\widehat\bPi_1\bA_\pm=\bA_\pm,	&&\widehat\bPi_2\bA_\pm=\widehat\bPi_1\bB_\pm=\bzero,
	&&\widehat\bPi_2\bB_\pm=\bB_\pm,
	\label{proj-AB}
	\end{align}
Eqs.~(\ref{complete-proj}) and (\ref{asym-condi-1}) imply
	\begin{align}
	&\bA_\pm+\bB_\pm=\bC_\pm,
	\label{AB=C-2}\\
	&\bcA_-=\bA_-,\quad\quad\quad \bcB_+=\bB_+,
	\end{align}
and we can use (\ref{commuting}), (\ref{Phi-t-asymp-pm}), and (\ref{cA-cB-pm}) to show that
	\be
	\tilde\bPhi(\cdot,z)\to e^{iz\widehat\varpi}\bA_\pm+
	e^{-iz\widehat\varpi}\bB_\pm=e^{-iz\widehat{\tilde\bH}_0}\bC_\pm
	~~~\for~~~z\to\pm\infty.
	\label{asym-condi-3}
	\ee
In particular, the $4$-component function defined by, $\bPsi(\cdot,z) :=e^{iz\widehat{\tilde\bH}_0}\tilde\bPhi(\cdot,z)$, satisfies
	\be
	\bPsi(\cdot,z)\to\bC_\pm~~~\for~~~z\to\pm\infty.
	\label{asym-condi-4}
	\ee

Because $\widehat{\tilde\bH}_0$ describes the propagation of waves in the absence of interactions, $\bPsi(\cdot,z)$ plays the role of the interaction-picture momentum-space wave functions \cite{Sakurai}. Expressing $\tilde\bPhi(\vec p,z)$ in terms of $\bPsi(\vec p,z)$ and substituting the result in (\ref{TD-sch-eq-FT}), we find
	\be
	i\partial_z\bPsi(\cdot,z)=\widehat\bcH\bPsi(\cdot,z)+\bfJ(\cdot,z),
	\label{TD-sch-eq-int}
	\ee
where
	\begin{align}
	&\widehat\bcH(z):=e^{iz\widehat{\tilde\bH}_0}
	(\widehat{\tilde\bH}-\widehat{\tilde\bH}_0)e^{-iz\widehat{\tilde\bH}_0},
	&&\bfJ(\cdot,z):=e^{iz\widehat{\tilde\bH}_0}\tilde\bJ(\cdot,z).
	\label{sH-fJ}
	\end{align}
We can express the general solution of the non-homogenous Schr\"odinger equation (\ref{TD-sch-eq-int}) in terms of the evolution operator for the corresponding homogeneous equation, namely the operator $\widehat\bcU(z,z_0)$ satisfying $i\partial_z \widehat\bcU(z,z_0)=\widehat\bcH(z)\widehat\bcU(z,z_0)$ and $ \widehat\bcU(z_0,z_0)=\widehat\bI$, for all $z,z_0\in\R$. This gives
	\be
	\bPsi(\cdot,z)=\widehat\bcU(z,z_0)\Big[
	\bPsi(\cdot,z_0)-i\int_{z_0}^z dz'\:
	\widehat\bcU(z_0,z')\bfJ(\cdot,z')\Big],
	\label{gen-sol-int}
	\ee
where we have made use of  $\widehat\bcU(z',z_0)^{-1}=\widehat\bcU(z_0,z')$. We also recall that $\widehat\bcU(z,z_0)$ admits the Dyson series expansion \cite{Sakurai}:
	\be
	\widehat\bcU(z,z_0)=\widehat\bI+\sum_{\ell=1}^\infty (-i)^\ell
        \int_{z_0}^z \!\!dz_\ell\int_{z_0}^{z_\ell} \!\!dz_{\ell-1}
        \cdots\int_{z_0}^{z_2} \!\!dz_1\,
        \widehat{\bcH}(z_\ell)\widehat{\bcH}(z_{\ell-1})\cdots\widehat{\bcH}(z_1).
        \label{dyson}
        \ee
       
Following Ref.~\cite{pra-2023}, we define the fundamental transfer matrix of the medium according to
	\begin{align}
	&\widehat\bM:=\Lpi\, \widehat\bcU(+\infty,-\infty)\, \Lpi.
	\label{M-def}
	\end{align}
This is a linear operator acting in $\sF^4_k:=\left\{\bF\in\sF^4\,|\, \bF(\vec p)=\bzero~~\for~~|\vec p|\geq k\,\right\}$. In view of (\ref{bABC-def}), (\ref{asym-condi-4}), (\ref{gen-sol-int}), and (\ref{M-def}), 
	\be
	\bC_+=\widehat\bM\,\bC_-+\bD,
	\label{C=C-J}
	\ee
where
	\be
	\bD:=-i\Lpi\int_{-\infty}^\infty dz'\:
	\widehat\bcU(+\infty,z')\bfJ(\cdot,z'),
	\label{D=}
	\ee
and we have benefitted from the identity $\widehat\bcU(+\infty,-\infty)\,\widehat\bcU(-\infty,z')=\widehat\bcU(+\infty,z')$.

\section{Radiation by an oscillating source in a linear medium}
\label{S3}

If the source is localized or more generally confined to the region bounded by the planes $z=a_\pm$, the emitted wave satisfies the out-going asymptotic boundary condition. To quantify this condition, we examine the behavior of the $4$-component fields $\bPhi$ for $z\to\pm\infty$. Performing the inverse Fourier transform of both sides of (\ref{asym-condi-3}), we find
	\be
	\bPhi(\vec r,z)\to\frac{1}{4\pi^2}\int_{\sD_k}d^2 p\: e^{i\vec p\cdot\vec r}\left[
	\bA_\pm(\vec p)e^{i\varpi(\vec p)z}+\bB_\pm(\vec p)e^{-i\varpi(\vec p)z}\right]~~\for~~z\to\pm\infty,
	\label{asym-condi-3-IFT}
	\ee
where $\sD_k:=\{\vec p\in\R^2\,|\,|\vec p|<k\}$. This identifies $\bA_\pm$ and $\bB_\pm$ with the Fourier coefficients of the right- and left-going waves along the $z$ axis, respectively. Therefore, the out-going asymptotic boundary condition corresponds to $\bA_-=\bB_+=\bzero$. In view of (\ref{AB=C-2}), we can also write it in the form $\bC_-=\bB_-$ and $\bC_+=\bA_+$.
Substituting these equations in (\ref{proj-AB}), we have
	\begin{align}
	&\widehat\bPi_2\bC_-=\bC_-,
	&&\widehat\bPi_1\bC_-=\widehat\bPi_2\bC_+=\bzero,
	&&\widehat\bPi_1\bC_+=\bC_+.
	\label{proj-CC}
	\end{align}
Next, we apply $\widehat\bPi_j$ to both sides of (\ref{C=C-J}) and use (\ref{proj-CC}) to establish
	\bea
	&&\widehat\bPi_2\widehat\bM\,\bC_-=-\widehat\bPi_2\bD,
	\label{C-minus=1}\\
	&&\bC_+=\widehat\bPi_1(\widehat\bM-\widehat\bI)
	\bC_-+\widehat\bPi_1\bD.
	\label{C-plus=1}
	\eea
These are linear equations for $\bC_\pm$ which are to be solved in $\sF^4_k$. 
	
Equations (\ref{C-minus=1}) and (\ref{C-plus=1}) have the same structure as Eqs.~(57) and (58) of Ref.~\cite{pra-2023} for the $4$-component fields $\bT^l_\pm$ that determine the asymptotic expression for the scattered waves. The only difference is that in the scattering set-up considered in \cite{pra-2023}, $\bD$ is determined by the incident wave whose source resides at $z=-\infty$. This suggests that we can pursue the approach of Refs.~\cite{pra-2023,jpa-2020} to express the scaled electric field $\bcE$ for the electromagnetic wave reaching the detectors placed at $z=\pm\infty$ as follows.
	\be
	\bcE(\bfr)=\frac{k|\cos\vartheta|e^{ikr}}{2\pi i \,r}\:
	\bXi(\vartheta,\varphi)^T\bC_\pm(\vec k)~~~~\for~~~	r\cos\vartheta\to\pm\infty,
	\label{cE-detected}
	\ee
where $r$, $\vartheta$, and $\varphi$ are respectively the radial, polar, and azimuthal spherical coordinates of the position $\bfr$ of the detector, 
	\begin{align}
	&\bXi(\vartheta,\varphi)^T:=\left[\!\begin{array}{cccc}
	\bfe_x &
	\bfe_y &
	\sin\vartheta\sin\varphi\,\bfe_z &
	-\sin\vartheta\cos\varphi\,\bfe_z\end{array}\!\right],
	\label{vec-k-bXi=}
	\end{align}
and $\vec k$ is the projection of the wave vector $\bk:=k\,\hat\bfr$ onto the $x$-$y$ plane, i.e.,
	\be
	\vec k:=\frac{k\,\vec r}{r}=k\,(\sin\vartheta\cos\varphi\,\bfe_x+\sin\vartheta\sin\varphi\,\bfe_y).
	\label{vec-k-def}
	\ee 
With the help (\ref{vec-k-bXi=}), we can show that 
	\be
	\bXi(\vartheta,\varphi)^T\bC_\pm(\vec k)=\bc_\pm^++[(\bc_\pm^-\times\hat\bfr)\cdot\bfe_z]\bfe_z ~~~\for~~~\pm\cos\vartheta>0,
	\label{Xi-Tpm=}
	\ee	
where 
	\begin{align}
	&\bc^+_\pm:=C_{\pm1}(\vec k)\,\bfe_x+C_{\pm2}(\vec k)\,\bfe_y,
	&&\bc^-_\pm:=C_{\pm3}(\vec k)\,\bfe_x+C_{\pm4}(\vec k)\,\bfe_y,
	\label{bcpm=}
	\end{align} 
and $C_{\pm i}$ are the entries of $\bC_\pm$, so that $\bC_\pm^T=\left[C_{\pm1}~~C_{\pm2}~~C_{\pm3}~~C_{\pm4}\right]$.  	
			
Equation~(\ref{cE-detected}) reduces the solution of the radiation problem for oscillating sources to the determination of the $4$-components functions $\bC_\pm$. 
The entries of these are however not indepenent. To see this, we introduce the two-component functions,
	\begin{align}
	&\vec C_\pm^+:=\left[\begin{array}{c}
	C_{\pm 1}\\
	C_{\pm 2}\end{array}\right],
	&&\vec C_\pm^-:=\left[\begin{array}{c}
	C_{\pm 3}\\
	C_{\pm 4}\end{array}\right],
	\label{C-pm-def}
	\end{align}
so that 
	\be
	\bC_\pm=\left[\begin{array}{c}
	\vec C_{\pm}^+\\
	\vec C_{\pm}^-\end{array}\right].
	\label{2-comp-Cpm}
	\ee
Employing (\ref{proj}), (\ref{proj-CC}), and (\ref{2-comp-Cpm}), we then find
	\be
	\vec C_\pm^+(\vec p)=\mp\frac{1}{\varpi(\vec p)}\,\tilde\bL_0(\vec p)\vec C_\pm^-(\vec p).
	\label{C-minus-identity}
	\ee
	
Next, we use (\ref{t-H-zero}), (\ref{vec-k-def}), (\ref{C-minus-identity}), and $\varpi(\vec k)=|k_z|$, to show that 
	\be
	\vec C_\pm^+(\vec k)=-\frac{k}{2|k_z|}\left[\!\begin{array}{cc}
	1-\sin^2\vartheta\cos^2\varphi & 
	-\sin^2\vartheta\sin\varphi\cos\varphi\\
    	-\sin^2\vartheta\sin\varphi\cos\varphi & 
	1-\sin^2\vartheta\sin^2\varphi
    	\end{array}\!\right]\vec G_\pm(\vec k),
	\label{C-minus-identity-2}
	\ee
where 
	\be
	\vec G_\pm(\vec p):=\mp 2i\,\bsigma_2\vec C^-_\pm(\vec p).
	\label{vG-pm-def}
	\ee 
Substituting (\ref{C-minus-identity-2}) in (\ref{C-pm-def}) to determine $C_{\pm i}$ and using the result in (\ref{bcpm=}), we have
	\be
	\bc^+_\pm=-\frac{k}{2|k_z|}\left[\bg_\pm-(\hat\bfr\cdot\bg_\pm)\hat\bfr+(\hat\bfr\cdot\bg_\pm)(\bfe_z\cdot\hat\bfr)\bfe_z\right],
	\label{bcpm=2}
	\ee
where $\bg_\pm$ is the vector lying in the $x$-$y$ plane whose $x$ and $y$ components respectively coincide with the first and second entries of $\vec G_\pm(\vec k)$, i.e.,
	\begin{align}
	\bg_\pm&:=\big\{[\begin{array}{cc}
	\!\!1 & 0\!\!\end{array}]\vec G_\pm(\vec k)\big\}\,\bfe_x+
	\big\{[\begin{array}{cc}
	\!\!0 & 1\!\!\end{array}]\vec G_\pm(\vec k)\big\}\,\bfe_y
	\label{gpm-def=0}\\
	&\:=\pm 2\, \bfe_z\times\bc^-_\pm.
	\label{gpm-def=1}
	\end{align}
The latter equation follows from (\ref{bcpm=}), (\ref{C-pm-def}), and (\ref{vG-pm-def}). Solving it for $\bc^-_\pm$, we find
	\be
	\bc^-_\pm=\pm \frac{1}{2}\:\bg_\pm\times\bfe_z.
	\label{bcpm=3}
	\ee
We can express $\bc^+_\pm$ by substituting (\ref{bcpm=3}) in (\ref{bcpm=2}). More interesting is the identity,
	\be
	\bc_\pm^++[(\bc_\pm^-\times\hat\bfr)\cdot\bfe_z]\bfe_z
	=\frac{k}{2|k_z|}\,\hat\bfr\times(\hat\bfr\times\bg_\pm)~~~\for~~~\pm\cos\vartheta>0,\nn
	\ee
which in view of (\ref{sE-sH}), (\ref{cE-detected}), and (\ref{Xi-Tpm=}) leads us to the following remarkably simple equation for the electric field of the wave reaching the detectors.
	\be
	\boldsymbol{\mathsf{E}}(\bfr,t)=\frac{k\,e^{i(kr-\omega t)}}{4\pi i\sqrt{\varepsilon_0}\,r}
	\:\,\hat\bfr\times(\hat\bfr\times\bg_\pm)~~~\for~~~r\cos\vartheta\to\pm\infty.
	\label{cE-detected-new}
	\ee
According to this equation, the information about the radiation of the source is contained in a pair of vectors lying in the $x$-$y$ plane, namely $\bg_\pm$.\footnote{This holds also for the scattering of electromagnetic waves where the roles of $\bC_\pm$ and $\bc^\pm_\pm$ are respectively played by $\bT_\pm^{l/r}$ and  $\bt^\pm_\pm$ of Ref.~\cite{pra-2023}. In particular, we can express the electric field of the scattered wave arriving at the detectors in terms of the vectors $\bg_\pm$ defined by (\ref{gpm-def=0}) with $\bc^\pm_\pm$ changed to $\bt^\pm_\pm$.}

	
Next, we explore the utility of (\ref{cE-detected-new}) in solving the textbook problem of the radiation of an oscillating source placed in vacuum \cite{Jackson}.

In the absence of scatterers, $\hat\bvarepsilon=\hat\bmu=\bI$, 
	\begin{align*}
	&\widehat{\tilde\bH}=\tilde\bH_0,
	&&\widehat{\sH}(z)=\widehat\bzero,
	&& \widehat\bcU(z,z')=\widehat\bI, 
	&&\widehat\bcM=\widehat\bI, 
	&&\widehat\bM=\Lpi,
	\end{align*} 
and Eqs.~(\ref{D=}), (\ref{C-minus=1}), and (\ref{C-plus=1}) respectively take the form
	\begin{align}
	&\bD=-i\Lpi\int_{-\infty}^\infty dz\:\bfJ(\cdot,z),
	&&\bC_-=-\widehat\bPi_2\bD,
	&&\bC_+=\widehat\bPi_1\bD.
	\label{DCC=3}
	\end{align}
In view of (\ref{proj-H0}), (\ref{op-Pi}), (\ref{project}), (\ref{sH-fJ}), and (\ref{DCC=3}), 
	\bea
	&&\bC_-(\vec p)=i\chi_k(\vec p)\int_{-\infty}^\infty dz\:
	e^{iz\varpi(\vec p)}\bPi_2(\vec p)\tilde\bJ(\vec p,z)=
	i\chi_k(\vec p)\bPi_2(\vec p)\vardbtilde\bJ(\vec p,-\varpi(\vec p)),
	\label{C-minus=4}\\
	&&\bC_+(\vec p)=-i\chi_k(\vec p)\int_{-\infty}^\infty dz\:
	e^{-iz\varpi(\vec p)}\bPi_1(\vec p)\tilde\bJ(\vec p,z)=
	-i\chi_k(\vec p)\bPi_1(\vec p)\vardbtilde{\bJ}(\vec p,\varpi(\vec p)),
	\label{C-plus=4}
	\eea
where
	\[\chi_k(\vec p):=\left\{\begin{array}{ccc}
	1 &\for& |\vec p|<k,\\
	0 &\for& |\vec p|\geq k,\end{array}\right.\]
$\vardbtilde{\bJ}$ stands for the three-dimensional Fourier transform of $\bJ$, i.e., 	
	\be
	\vardbtilde{\bJ}(\vec p,p_z):=\int_{-\infty}^\infty dz\:e^{-ip_z z}\tilde\bJ(\vec p)=\int_{\R^3}d^3\bfr\: e^{-i\bp\cdot\bfr} \bJ(\bfr),
	\label{3D-FT}
	\ee
and $\bp$ denotes $\vec p+p_z\bfe_z$ which we also express as $(\vec p,p_z)$. Recall that the wave vector $\bk$ for a detected wave is given by $\bk:=k\hat\bfr$ and that the detectors lie on the planes $z=\pm\infty$. In particular, $\bk\in\R^3$, $|\bk|=k>0$, and $k_z=k\cos\vartheta\neq 0$. These in turn imply $\chi_k(\vec k)=1$ and $\bk=\vec k\pm\varpi(\vec k)\,\bfe_z$ for $\pm\cos\vartheta>0$.
Using these relations together with (\ref{C-minus=4}) and (\ref{C-plus=4}), we obtain
	\begin{align}
	&\bC_-(\vec k)=i\bPi_2(\vec k)\,\vardbtilde{\bJ}(\bk)
	~~~~~\for~~~\cos\vartheta<0,
	\label{C-minus=5}\\
	&\bC_+(\vec k)=-i\bPi_1(\vec k)\,\vardbtilde{\bJ}(\bk)
	~~~\for~~~\cos\vartheta>0.
	\label{C-plus=5}
	\end{align}
	
Next, we introduce $\vec P:=[\!\begin{array}{cc}
	p_x &p_y\end{array}\!]^T$,
and use (\ref{H-J-def}) and (\ref{t-H-zero}) to show that
	\begin{align}
	&\tilde{\bJ}(\vec p,z)=\frac{i}{k}\left[\begin{array}{c}
	\tilde{\cJ}_z(\vec p, z)\vec P\\[3pt] 
	ik\bsigma_2\tilde{{\vec\cJ}}(\vec p, z)\end{array}\right],
	\label{ttJ-K}\\[3pt]
	&ik^{-1}\bsigma_2\tilde\bL_0(\vec p)\vec P=\vec P.
	\label{ttJ-K2}
	\end{align}
These equations together with (\ref{proj}) and (\ref{L2-id}) imply 
	\begin{align}
	\bPi_1(\vec p)\vardbtilde{\bJ}(\vec p,p_z)=
	\frac{1}{2\varpi(\vec p)}\left[\begin{array}{c}
	\tilde\bL_0(\vec p)\,\bsigma_2\,\vec\sG_-(\vec p,p_z)\\
	-\varpi(\vec p)\,\bsigma_2\,\vec\sG_-(\vec p,p_z)\end{array}\right],
	\label{Pi1J}\\
	\bPi_2(\vec p)\vardbtilde{\bJ}(\vec p,p_z)=
	-\frac{1}{2\varpi(\vec p)}\left[\begin{array}{c}
	\tilde\bL_0(\vec p)\,\bsigma_2\,\vec\sG_+(\vec p,p_z)\\
	\varpi(\vec p)\,\bsigma_2\,\vec\sG_+(\vec p,p_z)\end{array}\right],
	\label{Pi2J}
	\end{align}
where
	\bea
	\vec\sG_\pm(\vec p,p_z)&:=&\vardbtilde{\vec\cJ}(\vec p,p_z)\pm
	\varpi(\vec p)^{-1}\vardbtilde{{\cJ}}_z(\vec p,p_z)\vec P.
	\label{bg=pm}
	\eea
In particular, for $\bp=\bk$, we have $\vec p=\vec k$, $\vec P=\vec\fK:=[\!\begin{array}{cc}
	k_x &k_y\end{array}\!]^T=
	[\!\begin{array}{cc}
	k\sin\vartheta\cos\varphi &
	k\sin\vartheta\sin\varphi\end{array}\!]^T$, and
(\ref{C-minus=5}), (\ref{C-plus=5}), and (\ref{Pi1J}) -- (\ref{bg=pm}) yield
	\bea
	\bC_\pm(\vec k)&=&\frac{i}{2|k_z|}\left[\begin{array}{c}
	-\tilde\bL_0(\vec k)\,\bsigma_2\,\vec G_0(\bk)\\
	k_z\,\bsigma_2\,\vec G_0(\bk)\end{array}\right]~~~\for~~~\pm\cos\vartheta>0,
	\label{Cpm=-987}
	\eea
where
	\bea
	\vec G_0(\bk)&:=&\vardbtilde{{\vec\cJ}}(\bk)-
	k_z^{-1}\vardbtilde{{\cJ}}_z(\bk)\vec\fK
	=\frac{k}{k_z}\left[\begin{array}{c}
	\bfe_y\cdot[\hat\bfr\times\vardbtilde{{\!\bcJ}}(\bk)]\\
	-\bfe_x\cdot[\hat\bfr\times\vardbtilde{{\!\bcJ}}(\bk)]
	\end{array}\right],
	\label{bg=}
	\eea
and we have also benefitted from the fact that 
	\be
	k_z=\pm\varpi(\vec k)~~~\for~~~\pm\cos\vartheta>0.
	\label{fact-1}
	\ee

In view of (\ref{2-comp-Cpm}), (\ref{Cpm=-987}), and (\ref{fact-1}), $\vec C^-_\pm(\vec k)=\pm\frac{i}{2}\,\bsigma_2\vec G_0(\bk)$ for $\pm\cos\vartheta>0$. Using this relation in 
(\ref{vG-pm-def}), we find that for the system we consider,
	\[\vec G_\pm(\vec k)=\vec G_0(\bk)~~~\for~~~\pm\cos\vartheta>0.\]
This equation together with (\ref{gpm-def=0}) and (\ref{bg=}) imply
	\be
	\bg_\pm=\bg_0~~~\for~~~\pm\cos\vartheta>0,
	\label{bg-pm-vacuum}
	\ee
where 
	\bea
	\bg_0&:=&-\frac{k}{k_z}\,\bfe_z\times\big[\hat\bfr\times\vardbtilde{{\!\bcJ}}(\bk)\big]=\vardbtilde{{\!\bcJ}}(\bk)-\sec\vartheta\,\vardbtilde{\cJ}_z(\bk)\,\hat\bfr.
	\label{vec-g=}
	\eea
Substituting (\ref{bg-pm-vacuum}) in (\ref{cE-detected-new}) and making use of (\ref{rho-sJ}), (\ref{vec-g=}), and the fact that $\boldsymbol{\mathsf{E}}$ is a continuous function of $\hat\bfr$, we have	
	\bea
	\boldsymbol{\mathsf{E}}(\bfr,t)
	&=&\frac{k\,e^{i(kr-\omega t)}}{4\pi i\sqrt{\varepsilon_0}\,r}
	\:\,\hat\bfr\times[\hat\bfr\times\vardbtilde{{\!\bcJ}}(\bk)]~~~~~~~~~~\:\for~~~r\to\infty
	\nn\\
	&=&\frac{kZ_0}{4\pi i}\:\frac{e^{i(kr-\omega t)}}{r}\: 
	\hat\bfr\times[\hat\bfr\times
	\vardbtilde{{\boldsymbol{\mathsf{J}}}}(k\hat\bfr,0)] 
	~~~~\for~~~r\to\infty,
	\label{cE-detected-vacuum}
	\eea
where $Z_0:=\sqrt{\mu_0/\varepsilon_0}$ is the vacuum impedance, and $\vardbtilde{{\boldsymbol{\mathsf{J}}}}(k\hat\bfr,t):=\int_{\R^3}d^3r\:e^{-i\bk\cdot\bfr}
{\boldsymbol{\mathsf{J}}}(\bfr,t)$.

Equation (\ref{cE-detected-vacuum}) coincides with the outcome of the standard treatment of the radiation of an oscillating source in vacuum \cite{Jackson}; Eqs.~(9.4), (9.5), and (9.8) of Ref.~\cite{Jackson} also imply (\ref{cE-detected-vacuum}). Note however that our derivation of this equation is manifestly gauge-invariant; unlike its textbook derivation, it does not involve the retarded vector potential and the Lorentz gauge condition.

Although the derivation of (\ref{cE-detected-vacuum}) rests on the assumption that the source of the radiation resides in empty space, we can also use it to describe the radiation of the source in a general linear medium provided that we characterize the electromagnetic response of the medium not in terms of its permittivity and permeability tensors but its bound charge density and bound and polarization current densities \cite{griffiths-EM}. This requires replacing the free current density ${{\boldsymbol{\mathsf{J}}}}$ appearing in (\ref{cE-detected-vacuum}) with the sum of the free, bound, and polarization current densities; $
	{{\boldsymbol{\mathsf{J}}}}\to {{\boldsymbol{\mathsf{J}}}}':=
	\boldsymbol{\mathsf{J}}+\boldsymbol{\mathsf{J}}_{\rm b}+
	\boldsymbol{\mathsf{J}}_{\rm p}$,
where $\boldsymbol{\mathsf{J}}_{\rm b}$ and $\boldsymbol{\mathsf{J}}_{\rm p}$ respectively stand for the bound and polarization current densities. Because we do not know the explicit form of $\boldsymbol{\mathsf{J}}_{\rm b}$ and $\boldsymbol{\mathsf{J}}_{\rm p}$, the substitution of ${{\boldsymbol{\mathsf{J}}}}'$ for ${{\boldsymbol{\mathsf{J}}}}$ in (\ref{cE-detected-vacuum}) gives a formula for the electric field of the detected wave which is of little practical value. Nevertheless, using the description of the medium in terms of the effective current density ${{\boldsymbol{\mathsf{J}}}}'$ we can establish the general validity of Eq.~(\ref{bg-pm-vacuum}) provided that we let ${{\boldsymbol{\mathsf{J}}}}'$ play the role of $\boldsymbol{\mathsf{J}}$ in the definition of $\bg_0$. In other words, we have	\begin{align}
	&\bg_\pm=\bg~~~\for~~~\pm\cos\theta>0,
	\label{equal-bg}
	\end{align} 
where 
	\be
	\bg:=\vardbtilde{{\!\bcJ}}'(\bk)-\sec\vartheta\,\vardbtilde{\cJ}'_z(\bk)\,\hat\bfr,
	\label{bg-equal}
	\ee
$\vardbtilde{{\!\bcJ}}':=\vardbtilde{{\!\bcJ}}+
\vardbtilde{{\!\bcJ}}_{\rm b}+\vardbtilde{{\!\bcJ}}_{\rm p}$, ~${{\!\bcJ}}_{\rm b}$ and ~${{\!\bcJ}}_{\rm p}$ are the vector-valued functions satisfying $\boldsymbol{\mathsf{J}}_{\rm p}(\bfr,t)=\mu_0^{-1/2} e^{-i\omega t}\bcJ_{\rm p}(\bfr)$
and $\boldsymbol{\mathsf{J}}_{\rm b}(\bfr,t)=\mu_0^{-1/2} e^{-i\omega t}\bcJ_{\rm b}(\bfr)$, and $\vardbtilde{\cJ}'_z$ is the $z$ component of $\,\vardbtilde{\!\bcJ}'$.

Substituting (\ref{equal-bg}) in (\ref{cE-detected-new}) and making use of the fact that the electric field is a continuous function of $\vartheta$, we find the following expression for the electric field of the detected wave.
	\be
	\boldsymbol{\mathsf{E}}(\bfr,t)=\frac{k\,e^{i(kr-\omega t)}}{4\pi i\sqrt{\varepsilon_0}\,r}
	\:\,\hat\bfr\times(\hat\bfr\times\bg)~~~\for~~~r\to\infty.
	\label{cE-detected-new2}
	\ee
Because $\bg$ is orthogonal to the $z$ axis, the right-hand side of this relation is not manifestly covariant. The definition of $\bg$, however, shows that $\hat\bfr\times\bg=\hat\bfr\times\vardbtilde{\!\bcJ}'$. This confirms the covariance of $\hat\bfr\times\bg$ and consequently that of the right-hand side of (\ref{cE-detected-new2}). 

We would like to emphasize that because we do not know the explicit form of ${{\!\bcJ}}'$,  we cannot use (\ref{bg-equal}) to compute $\bg$. We can instead employ our transfer-matrix approach and the knowledge of the permittivity and permeability tensors of the medium to determine $\bg$.

This completes our general discussion of the radiation problem for an oscillating source in a linear medium. It leads to the following prescription for computing the electric field of the wave arriving at the detectors.
	\begin{enumerate}
	\item Find the evolution operator $\widehat\bcU(+\infty,z)$, the $4$-component field $\bD$, and the transfer matrix $\widehat\bM$.
	\item Solve (\ref{C-minus=1}) for $\bC_-$.
	\item Read off the expressions for $\vec C_-^-$, $\vec G_-$, and $\bg_-$ using (\ref{2-comp-Cpm}), (\ref{vG-pm-def}), and (\ref{gpm-def=0}).
	\item Substitute $\bg_-$ for the $\bg$ in (\ref{cE-detected-new2}).
	\end{enumerate}

\section{Radiation of an oscillating source in the presence of point scatterers}
\label{S4}

Suppose that an oscillating source is placed next to a finite collection of nonmagnetic point scatterers lying in the $x$-$y$ plane with otherwise arbitrary positions, so that the relative permittivity and permeability tensors of the system take the form
	\begin{align}
	&\hat\bvarepsilon(\vec r,z)=\bI+\delta(z)\sum_{a=1}^N\bfZ_a\,\delta(\vec r-\vec r_a),
	&&\hat\bmu\mbox{\normalsize$(\vec r,z)$}=\bI,
	\label{delta}
	\end{align}
where $N$ is the number of point scatterers, $\vec r_a=(x_a,y_a)$ signify their positions in the $x$-$y$ plane\footnote{$\vec r_a\neq\vec r_b$ for $a\neq b$.}, and $\bfZ_a$ are $3\times 3$ complex matrices with entries $\fZ_{a,ij}$. Then, as we show in Ref.~\cite{pra-2023}, for the generic cases where $\fZ_{a,33}\neq 0$, we find  the following expression for the interaction-picture Hamiltonian (\ref{sH-fJ}).
	\be
	\widehat\bcH(z)=i\delta(z)\widehat\bV,
	\label{pt-sH=}
	\ee
where
	\be
	\widehat\bV:=
	k\sum_{a=1}^N \left[\begin{array}{cc}
	\bzero&\bzero \\
	\bsZ_a&\bzero\end{array}\right]\widehat v_a,
	\label{bV=}
	\ee	
$\bsZ_a$ are the $2\times 2$ matrices given by
	\begin{align}
	\bsZ_a&:=\frac{1}{\fZ_{a,33}}\left[\begin{array}{cc}
	\fZ_{a,22}\fZ_{a,33}-\fZ_{a,23}\fZ_{a,32}&\fZ_{a,23}\fZ_{a,31}-\fZ_{a,21}\fZ_{a,33}\\
	\fZ_{a,13}\fZ_{a,32}-\fZ_{a,12}\fZ_{a,33}&\fZ_{a,11}\fZ_{a,33}-\fZ_{a,13}\fZ_{a,31}
	\end{array}\right]\bsigma_2,\nn\\
	&=\frac{1}{\fZ_{a,33}}\left[\begin{array}{cc}
	\cZ_{a,11}&-\cZ_{a,12}\\
	-\cZ_{a,21}&\cZ_{a,22}\end{array}\right]\bsigma_2=
	\frac{i}{\fZ_{a,33}}\left[\begin{array}{cc}
	-\cZ_{a,12}&-\cZ_{a,11}\\
	\cZ_{a,22}&\cZ_{a,21}\end{array}\right],
	\label{Z=}
	\end{align}
$\cZ_{a,ij}$ stands for the minor of the $\fZ_{a,ij}$ entry of $\bfZ_a$,  i.e., $\cZ_{a,ij}$ is the determinant of the $2\times 2$ matrix obtained by deleting the $i$-th row and $j$-th column of $\bfZ_a$, for every positive integer $m$, $\widehat v_a:\sF^m\to\sF^m$ is the linear operator\footnote{Ref.~\cite{pra-2023} uses $\tilde\delta(i\vec\nabla_p-\vec r_a)$ for what we call $\widehat v_a$.} defined by
	\be
	\widehat v_a\bF(\vec p):=\widecheck\bF(\vec r_a)\,e^{-i\vec r_a\cdot\vec p},
	\label{delta-op=}
	\ee
$\bF\in\sF^m$ is arbitrary, and $\widecheck\bF:=\cF^{-1}\bF$ is the two-dimensional inverse Fourier transform of $\bF$, i.e.,
$\widecheck\bF(\vec r):=\frac{1}{4\pi^2} \int_{\R^2} d^2 \vec p\,'\,e^{i\vec r_a\cdot\vec p\,'}\bF(\vec p\,')$.

According to (\ref{pt-sH=}), $\widehat\bcH(z_1)\widehat\bcH(z_2)=\bzero$ for all $z_1,z_2\in\R$. Therefore the Dyson series (\ref{dyson}) terminates, and we find 
	\be
	\widehat\bcU(+\infty,z)=\widehat\bI+\widehat\bV\int_{z}^\infty\delta(z')dz'=	
	\widehat\bI+\left\{\begin{array}{ccc}
	\widehat\bV &\for&z<0,\\
	\widehat\bzero &\for&z>0.
	\end{array}\right.
	\label{bcU-pt=}
	\ee
Notice that $\widehat\bcU(+\infty,0)$ involves $\int_0^\infty\delta(z')dz'$ which is ill-defined. This causes no problem in computing the fundamental transfer matrix (\ref{M-def}), because the latter requires the knowledge of $\,\widehat\bcU(+\infty,-\infty)$. According to (\ref{M-def}) and (\ref{bcU-pt=}), 
	\be
	\widehat\bM=\Lpi+\Lpi\widehat\bV\Lpi.
	\label{TM-point}
	\ee
In view of Eq.~(\ref{D=}), the problem with $\int_0^\infty\delta(z')dz'$ does not affect the calculation of the $4$-component function $\bD$ either, if $\tilde\bJ(\vec p,z)$  and consequently $\bfJ(\vec p,z)$ are continuous functions of $z$ at $z=0$. Under this condition, we can use (\ref{D=}) and (\ref{bcU-pt=}) to infer 
	\be
	\bD(\vec p)=-i\chi_k(\vec p)\left[\int_{-\infty}^\infty dz\:\bfJ(\vec p,z)
	+\int_{-\infty}^0 dz\: \widehat\bV\bfJ(\vec p,z)\right].
	\label{D=2}
	\ee
To obtain a more explicit expression for the right-hand side of this equation, we examine the structure of the $4$-components functions $\bfJ$ and $\bJ$ in the presence of the point scatterers given by (\ref{delta}). 

According to (\ref{H-J-def}), $\bJ$ involves $\hat\varepsilon_{33}^{-1}\cJ_z$ and $\hat\varepsilon_{33}^{-1}\cJ_z\vec K_\cE$. To deal with the fact that $\hat\varepsilon_{33}$ and $\vec K_\cE$ have delta-function singularities, we employ the following distributional identity which we prove in Appendix~A of Ref.~\cite{pra-2023}.
	\be
	\frac{\delta(\vec r-\vec r_a)}{1+\sum_{c=1}^N\fa_c\,\delta(\vec r-\vec r_c)}=0,\nn
	\ee
where $\fa_c$ are nonzero numbers. This together with (\ref{KEs=}) allow us to show that whenever $\cJ_z$ is a continuous function on the $x$-$y$ plane, $\hat\varepsilon_{33}^{-1}\cJ_z=\cJ_z$ and $\hat\varepsilon_{33}^{-1}\cJ_z\vec K_\cE=\bzero$. Substituting these in (\ref{H-J-def}) and taking note of (\ref{sH-fJ}), we see that the presence of point scatterers do not affect $\bJ$ or $\bfJ$. In particular, (\ref{ttJ-K}), (\ref{Pi1J}), and (\ref{Pi2J}) hold. In view of these equations and (\ref{exp=proj}) and (\ref{sH-fJ}), we have
	\bea
	\chi_k(\vec p)\int_{-\infty}^\infty dz\,
	\bfJ(\vec p,z)&=&\chi_k(\vec p)\left[\bPi_1(\vec p)\,\vardbtilde{\bJ}(\vec p,\varpi(\vec p))+
	\bPi_2(\vec p)\,\vardbtilde{\bJ}(\vec p,-\varpi(\vec p))\right]\nn\\
	&=&\frac{\chi_k(\vec p)}{2\varpi(\vec p)}
	\left[\begin{array}{cc}
	\tilde\bL_0(\vec p)\,\bsigma_2[\vec\sG_-(\vec p,\varpi(\vec p))-
	\vec\sG_+(\vec p,-\varpi(\vec p))]\\[3pt]
	-\varpi(\vec p)\,\bsigma_2[\vec\sG_-(\vec p,\varpi(\vec p))+
	\vec\sG_+(\vec p,-\varpi(\vec p))]\end{array}\right],
	\label{eq-32-1}
	\eea
where $\vec\sG_\pm$ are given by (\ref{bg=pm}).

Next, we use (\ref{t-H-zero}), (\ref{exp=2}), (\ref{sH-fJ}), (\ref{ttJ-K}), (\ref{bV=}), and (\ref{delta-op=}) to show that
	\be
	\chi_k(\vec p)\int_{-\infty}^0 dz\:\widehat\bV\bfJ(\vec p,z)=	
	\left[\begin{array}{c}
	\bzero\\[3pt] 
	\chi_k(\vec p)\sum_{a=1}^N e^{-i\vec r_a\cdot\vec p}\,\bsZ_a
	(\vec\fR_a+\vec\fS_a)\end{array}\right],
	\label{VJ=3}
	\ee
where 
	\bea
	\vec\fR_a&:=&k\int_{-\infty}^0 dz\:\vec\cR(\vec r_a,z),
	\quad\quad\quad
	\vec\fS_a:=k\int_{-\infty}^0 dz\:\vec\cS(\vec r_a,z),
	\label{fCS-def}\\
	\vec\cR(\vec r,z)&:=&\frac{i}{4\pi^2k}\int_{\sD_k}d^2 p\, 
	e^{i\vec p\cdot\vec r}\cos[z\varpi(\vec p)]\,\tilde\cJ_z(\vec p,z)\,\vec P,
	\label{fC-def}\\
	\vec\cS(\vec r,z)&:=&-\frac{i}{4\pi^2}\int_{\sD_k}d^2 p\; 
	\frac{e^{i\vec p\cdot\vec r}\sin[z\varpi(\vec p)]}{\varpi(\vec p)}\;
	\tilde\bL_0(\vec p)\bsigma_2\,\tilde{\vec \cJ}(\vec p,z).
	\label{fS-def}
	\eea
Substituting (\ref{eq-32-1}) and (\ref{VJ=3}) in (\ref{D=2}) we find the explicit form of the $4$-component function $\bD$. 
	
Let us recall that to determine the $4$-component functions $\bC_-$, we need to solve (\ref{C-minus=1}). First, we evaluate both sides of (\ref{C-minus=1}) 
at some $\vec p\in\R^2$ and use (\ref{TM-point}) to write the resulting equation in the form
	\bea
	\bC_-(\vec p)&=&-\bPi_2(\vec p)\left[\chi_k(\vec p)\widehat\bV\bC_-(\vec p)+\bD(\vec p)\right].
	\label{Cminus-pt=}
	\eea
We also note that (\ref{2-comp-Cpm}), (\ref{C-minus-identity}), (\ref{bV=}), and (\ref{delta-op=}) imply
	\begin{align}
	&\chi_k(\vec p)\widehat\bV\bC_-(\vec p)=k\left[\begin{array}{c}
	\bzero\\
	\vec X(\vec p)\end{array}\right],
	\label{eq-S32-302}
	\end{align}
where	
	\bea
	\vec X(\vec p)&:=&\chi_k(\vec p)\sum_{b=1}^N e^{-i\vec r_b\cdot\vec p}\:
	\bsZ_b\, \vec X_b,
	\label{X=}\\
	\vec X_b&:=&\widecheck{\,\vec C_-^+}(\vec r_b)=\frac{1}{4\pi^2}\int_{\R^2}d^2 p\:
	e^{i\vec p\cdot\vec r_b}\vec C_-^+(\vec p)=
	\frac{1}{4\pi^2}\int_{\R^2}d^2 p\:
	\frac{e^{i\vec p\cdot\vec r_b}}{\varpi(\vec p)}\tilde\bL_0(\vec p)\vec C_-^-(\vec p).
	\label{vx=}
	\eea
Next, we use the entries of $\bD$, which we denote by $D_i$, to introduce the $2$-component functions, $\vec D^+:=\left[\!\begin{array}{cc}
	D_{1} & D_{2}\end{array}\!\right]^T$ and $\vec D^-:=\left[\!\begin{array}{cc}
	D_{3} & D_{4}\end{array}\!\right]^T$, so that 
	\begin{align}
	&\bD:=\left[\begin{array}{c}
	\vec D^+\\
	\vec D^-\end{array}\right].
	\label{2-component}
	\end{align}
Substituting (\ref{eq-S32-302}) in (\ref{Cminus-pt=}) and using (\ref{proj}), (\ref{2-comp-Cpm}), 
(\ref{X=}), and (\ref{2-component}), we obtain
	\bea
	\vec C_-^-(\vec p)
	&=&-\frac{k\chi_k(\vec p)}{2}\sum_{a=1}^N 
	e^{-i\vec r_a\cdot\vec p}\bsZ_a\vec X_a+\vec\Delta(\vec p),
	\label{Cpm-m=xa}
	\eea
where  
	\begin{align}
	\vec\Delta(\vec p)
	&:=\frac{1}{2}\left[\frac{1}{\varpi(\vec p)}\,\tilde\bL_0(\vec p)\vec D^+(\vec p)-\vec D^-(\vec p)\right]\nn\\
	&\,=\frac{i\chi_k(\vec p)}{2}\Big[
	\sum_{a=1}^N e^{-i\vec r_a\cdot\vec p}\,\bsZ_a
	(\vec\fR_a+\vec\fS_a)-\bsigma_2\Big\{\,\vardbtilde{{\!\vec\cJ}}(\vec p,-\varpi(\vec p))+\frac{\vardbtilde{{\!\cJ}}_z
	(\vec p,-\varpi(\vec p))}{\varpi(\vec p)}\,\vec P\Big\}\Big].
	\label{Delta-mm-pp=}
	\end{align}
and we have also employed (\ref{L2-id}), (\ref{bg=pm}), and (\ref{D=2}) -- (\ref{VJ=3}).

Plugging (\ref{Delta-mm-pp=}) in (\ref{Cpm-m=xa}), setting $\vec p=\vec k$, and using (\ref{vG-pm-def}), we find
	\be
	\vec G_-(\vec k):=\vec G_0(\bk)+\vec G_s(\bk)~~~\for~~~\cos\vartheta<0,
	\label{bg=bg+bgs}
	\ee
where $\vec G_0(\bk)$ is given by (\ref{bg=}) and
	\be
	\vec G_s(\bk):=-\bsigma_2\sum_{a=1}^N e^{-i\vec r_a\cdot\vec k}\bsZ_a(\vec\fR_a+\vec\fS_a+ik\vec X_a).
	\label{bg-s-def}
	\ee
In view of  (\ref{gpm-def=0}), (\ref{bg=}), (\ref{vec-g=}), (\ref{equal-bg}), and (\ref{bg=bg+bgs}), $\bg=\bg_-=\bg_0+\bg_s$, where\footnote{We have also computed $\bC_+$, $\vec G_+$, and $\bg_+$ and checked that indeed $\bg_+=\bg_-$.}
	\be
	\bg_s:=\left\{[\begin{array}{cc}
	1 & 0\end{array}]\,\vec G_s(\bk)\right\}\bfe_x+\left\{[\begin{array}{cc}
	0 & 1\end{array}]\,\vec G_s(\bk)\right\}\bfe_y.
	\label{bgs=3}
	\ee
Susbtituting $\bg=\bg_0+\bg_s$ in (\ref{cE-detected-new2}) and using (\ref{vec-g=}), we have
	\begin{align}
	\boldsymbol{\mathsf{E}}(\bfr,t)&=\frac{k\,e^{i(kr-\omega t)}}{4\pi i\sqrt{\varepsilon_0}\,r}
	\:\left\{\hat\bfr\times[\hat\bfr\times\vardbtilde{{\!\bcJ}}(k\hat\bfr)]+
	\hat\bfr\times(\hat\bfr\times\bg_s)\right\}~~~\for~~~r\to\infty.
	\label{cE-detected-s}
	\end{align}

Equations (\ref{bg-s-def}) -- (\ref{cE-detected-s}) reduce the solution of the radiation problem we are considering to the determination of $\vec X_a$. To calculate the latter, first we use  (\ref{C-minus-identity}) and (\ref{Cpm-m=xa}) to derive
	\begin{align}
	&\vec C_-^+(\vec p)=-\frac{k\chi_k(\vec p)}{2}\sum_{a=1}^N 
	\frac{e^{-i\vec r_a\cdot\vec p}}{\varpi(\vec p)}\,\tilde\bL_0(\vec p)
	\bsZ_a\vec X_a+\vec\Gamma(\vec p),
	\label{Cpm-p=xa}
	\end{align}
where
	\be
	\vec\Gamma(\vec p):=\varpi(\vec p)^{-1}\tilde\bL_0(\vec p)\vec\Delta(\vec p).
	\label{Gamma-pm-def}
	\ee	
Performing the inverse Fourier transform of both sides of (\ref{Cpm-p=xa}), we then find 
	\be
	\widecheck{\vec C_-^+}(\vec r)=-\sum_{b=1}^N
	\bcL(\vec r-\vec r_b)\bsZ_b\vec X_b+\widecheck{\vec\Gamma}(\vec r),
	\label{eq-S32-303}
	\ee
where
	\begin{align}	
	\bcL(\vec r)&:=\frac{k}{8\pi^2}\int_{\sD_k}d^2 p\:
	\frac{e^{i\vec p\cdot \vec r}}{\varpi(\vec p)}\,\tilde\bL_0(\vec p).
	\label{cL-def}
	\end{align}
For $\vec r=\vec r_a$, (\ref{eq-S32-303}) gives the following system of equations for $\vec X_b$.
	\be
	\sum_{b=1}^N \bA_{ab}\,\vec X_b=\widecheck{\vec\Gamma}(\vec r_a),
	\label{system}
	\ee
where for all $a,b\in\{1,2,\cdots,N\}$,
	\begin{align}
	\bA_{ab}&:=\delta_{ab}\bI+\bcL_{ab}\bsZ_b,
	\label{Aab-ba=}\\
	\bcL_{ab}&:=\bcL(\vec r_a-\vec r_b).
	\label{bcL-ab=}
	\end{align}
	
Equation (\ref{system}) has a unique solution if and only if, for all $a,c\in\{1,2,\cdots,N\}$, there are $2\times 2$ matrices $\bB_{ac}$ such that 
	\be
	\sum_{b=1}^N\bB_{ab}\bA_{bc}=\delta_{ac}\bI.
	\label{Bab-def}
	\ee
Multiplying both sides of (\ref{system}) by $\bB_{ca}$ from the left, summing over $a$, and making use of (\ref{Bab-def}), we have 
	\be
	\vec X_c=\sum_{b=1}^N\bB_{cb}\widecheck{\vec\Gamma}(\vec r_b).
	\label{xc=}
	\ee
Substituting this equation in (\ref{bg-s-def}) and using the result in (\ref{cE-detected-s}), we obtain the electric field of the wave reaching the detectors.

\section{Radiation of a perfect dipole in the presence of point scatterers}
\label{S5}

By definition, the electric dipole moment of a charge distribution with charge density $\rho$ is given by $\bfp(t):=\int_{\R^3}d^3r\:\bfr \rho(\bfr,t)$. For a charge distribution corresponding to a  localized oscillating source of angular frequency $\omega$, we can use the continuity equation, $\bnabla\cdot \boldsymbol{\mathsf{J}}=i\omega\rho$, the identity, $\int_{\R^3}d^3r(\bnabla\cdot \boldsymbol{\mathsf{J}})\bfr=-\int_{\R^3}d^3r\:\boldsymbol{\mathsf{J}}$, and Eq.~(\ref{rho-sJ}), to show that \cite{Jackson}
	\be
	\bfp(t)=i\omega^{-1}
	\int_{\R^3}d^3r\:\boldsymbol{\mathsf{J}}(\bfr,t)=
	\frac{i\sqrt{\varepsilon_0}\,e^{-i\omega t}}{k} \int_{\R^3}d^3r\:
	\bcJ(\bfr).
	\ee
This suggests that we can model a perfect electric dipole by a scaled current density of the form
	\be
	\bcJ(\bfr)=\bj\,\delta(\bfr-\bfa),
	\label{dipole}
	\ee
where $\bj:=\frac{-ik}{\sqrt\varepsilon_0}\boldsymbol{\bfp}(0)$, $\boldsymbol{\bfp}(0)$ is the electric dipole moment of the dipole at $t=0$, and $\bfa$ is its position. 

According to (\ref{dipole}),
	\begin{align}
	&\tilde\bcJ(\vec p,z)=\bj\, \delta(z-a_z)e^{-i\vec a\cdot\vec p},
	&&\vardbtilde{{\!\bcJ}}(\bp)
	=\bj\,e^{-i\bfa\cdot\bp},
	\label{dipole2}
	\end{align}
where $\vec a:=(a_x,a_y)$ and $a_x,a_y,a_z$ are Cartesian components of $\bfa$, so that $\bfa=(a_x,a_y,a_z)=(\vec a,a_z)$. The first relation in (\ref{dipole2}) shows that $\tilde\bcJ(\vec p,z)$ is a continuous function of $z$ at $z=0$, and we can safely employ the results of the preceding section provided that $a_z\neq 0$, i.e., the dipole does not lie on the $x$-$y$ plane. In the following we assume that this condition holds. 

Using the second relation in (\ref{dipole2}) we can write (\ref{cE-detected-s}) in the form
	\be
	\boldsymbol{\mathsf{E}}(\bfr,t)
	=\frac{k\,e^{i(kr-\omega t)}}{4\pi i\sqrt{\varepsilon_0}\,r}\: 
	\Big[e^{-ik\bfa\cdot\hat r}\hat\bfr\times(\hat\bfr\times \bj)+
	\hat\bfr\times(\hat\bfr\times\bg_s)\Big]~~~\for~~~r\to\infty.
	\label{cE-detected-s-dipole}
	\ee
Therefore, to determine the electric field reaching the detectors we need to calculate $\bg_s$. To do this, first we use (\ref{fCS-def}) -- (\ref{fS-def}), (\ref{Delta-mm-pp=}), and (\ref{dipole2}) to calculate $\vec\fR_b$, $\vec\fS_b$, and $\vec\Delta$. This gives
	\begin{align}
	&\vec\fR_b=k j_z\theta(-a_z)\vec R_b(a_z) ,\quad\quad\quad\quad
	\vec\fS_b=k\,\theta(-a_z)\bS_b(a_z) \vec J,
	\label{fC-fS=dipole}\\
	&\begin{aligned}
	\vec\Delta(\vec p)=&\frac{i\chi_k(\vec p)}{2}\Big\{
	k\,\theta(-a_z)\sum_{b=1}^N e^{-i\vec r_b\cdot \vec p}\,\bsZ_b
	\big[j_z\,\vec R_b(a_z)+\bS_b(a_z)\vec J\:\big]\\
	&\hspace{1.8cm}- e^{-i\vec a\cdot\vec p} e^{ia_z\varpi(\vec p)}
	\bsigma_2\big[\vec J+\frac{j_z}{\varpi(\vec p)}\,\vec P\,\big]\Big\},
	\end{aligned}
	\label{Delta-mm-pp=dipole}
	\end{align}
where
	\begin{align}
	&\theta(x):=\left\{\begin{array}{cc}
	1 &\for~~x\geq 0,\\
	0 &\for~~x<0,\end{array}\right.
	\quad\quad\quad 
	\vec J:=\left[\begin{array}{c}j_x\\ j_y\end{array}\right],\\
	&\vec R_b(a_z):=\vec R(\vec r_b-\vec a,a_z),
	\quad\quad\quad\quad
	\bS_b(a_z):=\bS(\vec r_b-\vec a,a_z),
	\label{cSb=def}\\
	&\vec R(\vec r,z):=\frac{i}{4\pi^2k}\int_{\sD_k}d^2 p\; e^{i\vec r\cdot\vec p}\cos[z\varpi(\vec p)]\vec P,
	\label{cb=def}\\
	&\bS(\vec r,z):=-\frac{i}{4\pi^2}\int_{\sD_k}d^2 p\; 
	\frac{e^{i\vec r\cdot\vec p}
	\sin[z\varpi(\vec p)]}{\varpi(\vec p)}\:\tilde\bL_0(\vec p)\bsigma_2,
	\label{bS=def}
	\end{align}
and $j_x, j_y$, and $j_z$ are components of $\bj$.
	 
Next, we substitute (\ref{Delta-mm-pp=dipole}) in (\ref{Gamma-pm-def}) to find $\vec\Gamma(\vec p)$, take the inverse Fourier transform of the resulting equation, and use (\ref{L2-id}), (\ref{ttJ-K2}), (\ref{xc=}), and the identity, $\sum_{c=1}^N\bB_{ac}\bcL_{cb}\bsZ_b=\delta_{ab}\bI-\bB_{ab}$, which follows from (\ref{Aab-ba=}) and (\ref{Bab-def}), to show that
	\begin{align}
	\vec X_a
	=&\frac{i}{2}\sum_{b=1}^N\bB_{ab}\big[j_z\vec\sR_b(a_z)+\bsS_b(a_z)\vec J\big]+\nn\\
	&\hspace{.2cm}i\theta(-a_z)\Big\{\sum_{b=1}^N-\bB_{ab}\big[j_z\vec R_b(a_z)+
	\bS_b(a_z)\vec J\big]+j_z\vec R_a(a_z)+\bS_a(a_z)\vec J\Big\},
	\label{vec-Xa=}
	\end{align} 
where 	
	\begin{align}
	&\vec\sR_b(a_z):=\vec\sR(\vec r_b-\vec a,a_z),\quad\quad\quad
	\bsS_b(a_z):=\bsS(\vec r_b-\vec a,a_z),
	\label{db-Tb-def}\\
	&\vec\sR(\vec r,z):=\frac{i}{4\pi^2k}\int_{\sD_k}d^2 p\; 
	e^{i\vec r \cdot \vec p}e^{iz\varpi(\vec p)}\vec P,
	\label{da=def}\\
	&\bsS(\vec r,z):=-\frac{1}{4\pi^2}\int_{\sD_k}d^2 p\; 
	\frac{e^{i\vec r\cdot\vec p}e^{iz\varpi(\vec p)}}{\varpi(\vec p)}\,
	\tilde\bL_0(\vec p)\bsigma_2.
	\label{Ta=def}
	\end{align}
	
Notice that changing the term $e^{i\vec r \cdot \vec p}$ on the right-hand side of (\ref{cb=def}), (\ref{bS=def}), (\ref{da=def}) and (\ref{Ta=def})  to $e^{-i\vec r \cdot \vec p}$ multiplies the left-hand side of (\ref{cb=def}) and (\ref{da=def}) by a minus sign while not affecting the left-hand side of (\ref{bS=def}) and (\ref{Ta=def}). This shows that we can replace the $e^{i\vec r\cdot\vec p}$ in (\ref{cb=def}) and (\ref{da=def}) by $i\sin(\vec r\cdot\vec p)$, and in (\ref{bS=def}) and (\ref{Ta=def}) by $\cos(\vec r\cdot\vec p)$. Doing this, we find
	\begin{align}
	&\vec R(\vec r,z)=\RE[\vec\sR(\vec r,z)], &&\bS(\vec r,z)=\RE[\bsS(\vec r,z)],
	\label{cd-ST}
	\end{align}
where ``$\RE$'' stands for the real part of its argument. It is also worth mentioning that we can express $\vec\sR(\vec r,z)$ and $\bsS(\vec r,z)$ in terms of the function,
	\be
	h(\vec r,z):=\frac{1}{4\pi^2 k}\int_{\sD_k}d^2 p\: \frac{e^{i\vec r\cdot\vec p}
	e^{iz\sqrt{k^2-p^2}}}{\sqrt{k^2-p^2}}=
	\frac{1}{2\pi}\int_0^1 du\: \frac{u\, e^{ikz\sqrt{1-u^2}}J_0(k|\vec r| u)}{\sqrt{1-u^2}},
	\label{h-def}
	\ee
where $J_0$ stands for the zero-order Bessel function of the first kind.
It is easy to check that
	\bea
	\vec\sR(\vec r,z)&=&-i\vec\partial\;\partial_{z} h(\vec r,z),
	\label{R=h}\\
	\bsS(\vec r,z)&=&i\left[\begin{array}{cc}
	\partial_x^2+k^2 &\partial_x\partial_y\\
	\partial_x\partial_y&\partial_y^2+k^2\end{array}\right] h(\vec r,z).
	\label{S=h}
	\eea
	
According to (\ref{h-def}), the real and imaginary parts of $h(\vec r,z)$ are respectively even and odd functions of $z$. In view of (\ref{R=h}) and (\ref{S=h}), this implies that the real part of $\vec\sR(\vec r,z)$ and the imaginary part of $\bsS(\vec r,z)$ are even functions of $z$ while the imaginary part of $\vec\sR(\vec r,z)$ and the real part of $\bsS(\vec r,z)$ are odd functions of $z$. We can use these observations together with (\ref{cSb=def}), 
(\ref{db-Tb-def}), and (\ref{cd-ST}) to establish the following identities.
	\begin{align}
	&\vec R_b(-a_z)=\vec R_b(a_z),
	&&\bS_b(-a_z)=-\bS_b(a_z),
	\label{parity1}\\
	&\IM\big(\vec\sR_b(-a_z)\big)=-\IM\big(\vec\sR_b(a_z)\big),
	&&\IM\big(\bsS_b(-a_z)\big)=\IM\big(\bsS_b(a_z)\big),
	\label{parity2}
	\end{align}
where ``$\IM$'' stands for the imaginary part of its argument.	
	
Having calculated $\vec X_a$, we can use (\ref{bg-s-def}) and (\ref{fC-fS=dipole}) to establish
	\begin{align}
	\vec G_s(\bk)&=\frac{k}{2}\,\bsigma_2\!\!\sum_{a,b=1}^N e^{-i\vec r_a\cdot\vec k}\bsZ_a
	\bB_{ab}\Big\{j_z\vec\sR_b(a_z)+\bsS_b(a_z)\vec J-2\theta(-a_z)
	\big[j_z\vec R_b(a_z)+\bS_b(a_z)\vec J\,\big]\Big\}\nn
	\\
	&=\frac{k}{2}\,\bsigma_2\!\!\sum_{a,b=1}^N e^{-i\vec r_a\cdot\vec k}\bsZ_a
	\bB_{ab}\Big\{j_z\big[{\rm sgn}(a_z)\vec R_b(a_z)+i\IM\big(\vec\sR_b(a_z)\big)\big]+\nn\\
	&\hspace{5cm}\big[{\rm sgn}(a_z)\bS_b(a_z)+i\IM\big(\bsS_b(a_z)\big)\big]\vec J\Big\}\nn\\	
	&=\frac{k}{2}\,\bsigma_2\!\!\sum_{a,b=1}^N e^{-i\vec r_a\cdot\vec k}\bsZ_a
	\bB_{ab}\Big[{\rm sgn}(a_z)j_z\,\vec\sR_b(|a_z|)+\bsS_b(|a_z|)\vec J\Big],
	\label{vbgs=dipole2}
	\end{align}
where ${\rm sgn}(x):=x/|x|$ stands for the sign of $x$, and we have also made use of (\ref{cd-ST}), (\ref{parity1}), and (\ref{parity2}). Recalling that $\bg_s$ is the vector having the entries of $\vec G_s(\bk)$ as its components, we can read off the latter from (\ref{vbgs=dipole2}) and substitute the result in (\ref{cE-detected-s-dipole}) to obtain the electric field of the wave reaching the detectors. 

The appearance of ${\rm sgn}(a_z)$ on the right-hand side of (\ref{vbgs=dipole2}) might give the impression that it depends on our choice of the direction of the $z$ axis. This is unacceptable because $\bg_s$ enters the expression for the electric field which must not depend on our choice of the coordinate system. As a consistency check on the validity of (\ref{vbgs=dipole2}), we examine the behavior of its right-hand side under the change of coordinates $\fT$ that flips the sign of the $z$ component of all vectors. First we use (\ref{delta}) to infer that under this coordinate transformation, $\fZ_{a,ij}$ is left invariant unless one and only one of $i$ and $j$ is 3, in which case it changes sign. In light of (\ref{Z=}), (\ref{Aab-ba=}), (\ref{Bab-def}), this implies that $\bsZ_a$, $\bA_{ab}$, and consequently $\bB_{ab}$ are left invariant under $\fT$.  Equations (\ref{db-Tb-def}), (\ref{da=def}), and (\ref{Ta=def}) show that the same applies to $\vec\sR_b(|a_z|)$ and $\bsS_b(|a_z|)$. It is also clear that $\fT$ implies ${\rm sgn}(a_z)\to-{\rm sgn}(a_z)$, $j_z\to-j_z$, and $\vec J\to\vec J$. These observations prove the invariance of the right-hand side of (\ref{vbgs=dipole2}) under $\fT$.


In the remainder of this section we explore the consequences of our findings for the simplest special case, namely an oscillating perfect dipole in the presence of a single point scatterer ($N=1$).  Without loss of generality we choose our coordinate system in such a way that the point scatterer lies at the origin while the dipole is on the $z$ axis. Then, $\vec r_1=\vec a=\vec 0$, and (\ref{bgs=3}), (\ref{cL-def}), (\ref{Aab-ba=}) -- (\ref{Bab-def}), (\ref{db-Tb-def}) --  (\ref{cd-ST}), and (\ref{vbgs=dipole2}) imply
	\begin{align}
	&
	\bcL_{11}=\bcL(\vec 0)=-\frac{ik^3}{6\pi}\,\bsigma_2,
	\label{bcL-11=}\\
	&\bB_{11}=\bA_{11}^{-1}=\bfB:=\Big(\bI-\frac{ik^3}{6\pi}\,\bsigma_2\bsZ_1\Big)^{\!-1},
	\label{bB-11=}\\
	&
	\vec\sR_1(z)=\big[\!\begin{array}{cc} 0 & 0 \end{array}\!\big]^T, 
	\quad\quad\quad
	\bsS_1(z)=\frac{ik^2}{3\pi}\,\fs(kz)\,\bI,
	\label{bT-1=}\\
	&\vec G_s(\bk)=\fs(|a_z|k)(\bfB-\bI)\vec J,\\[3pt]
	&\bg_s=\fs(|a_z|k)\left[\big\{(\fB_{11}-1)j_x+\fB_{12}j_y\big\}\bfe_x+
	\big\{\fB_{21}j_x+(\fB_{22}-1)j_y\big\}\bfe_y\right],
	\label{bgs=1}
	\end{align}
where
	\bea
	&&\fs(x):=\frac{-3i[(x^2+ix-1)e^{ix}-\frac{x^2}{2}+1]}{2x^3}.
	\label{fs-def}
	\eea
The following are consequences of Eqs.~(\ref{delta}), (\ref{Z=}), (\ref{cE-detected-s}), (\ref{bB-11=}), (\ref{bgs=1}), and (\ref{fs-def}).
	\begin{enumerate}
	\item $\bg_s$ and consequently the electric field of the detected radiation do not depend on $j_z$ and blow up when $\det(\bI-\frac{ik^3}{6\pi}\,\bsigma_2\bsZ_1)=0$. This condition marks a spectral singularity of the system which can exist if the point scatterer is made of gain material \cite{prl-2009,Longhi-2010}. In the absence of a source, the spectral singularity corresponds to a configuration where the scatterer begins amplifying the background noise and emits coherent radiation \cite{pra-2011a,jo-2017}. This is the basic mechanism that applies to every laser. In the presence of a source, tuning the parameters of the system to approach to that a spectral singularity, so that $\det(\bI-\frac{ik^3}{6\pi}\,\bsigma_2\bsZ_1)\approx 0$, causes the point scatterer to function as an amplifier for the radiated wave.\footnote{Making $|\det(\bI-\frac{ik^3}{6\pi}\,\bsigma_2\bsZ_1)|$ too small leads to the emergence of nonlinear effects which renders our analysis inapplicable.}
	\item The term $\fs(|a_z|k)$ in (\ref{bgs=1}) determines the dependence of the electric field of the emitted wave on the distance $|a_z|$ between the dipole and the point scatterer. For $|a_z| \to 0$, $\fs(|a_z| k)\to 1$ and $\bg_s$ tends to a nonzero constant value. For $|a_z| k\to\infty$, $\fs(|a_z| k)\to 0$ and $\bg_s$ tends to zero. Therefore, as expected, the presence of a distant point scatterer ($|a_z|\gg k^{-1}$) does not have a noticeable affect on the radiation of the dipole. 
	\item The $z$ component of $\bj$ does not affect the response of the point scatterer to the radiation emitted by the dipole. This has to do with the fact that $\vec r_1-\vec a=\vec 0$ which causes $\vec\sR_1(z)$ to vanish for all $z$.
	\item If the principal axes of the point scatterer are aligned along the coordinate axes, there are $\fz_x,\fz_y,\fz_z\in\C$ such that
	\begin{align}
	&\bfZ_1=\left[\begin{array}{ccc}
	\fz_x & 0 & 0\\
	0 & \fz_y & 0\\
	0 & 0 & \fz_z\end{array}\right],
	&&\bsigma_2\bsZ_1=\left[\begin{array}{cc}
	\fz_x & 0\\
	0 & \fz_y\end{array}\right],
	&&
	\bfB=\left[\begin{array}{cc}
	\beta(\fz_x) & 0\\
	0 & \beta(\fz_y)\end{array}\right],\nn
	\end{align}
where 
	\be
	\beta(\fz):=\left(1-\frac{i\fz k^3}{6\pi}\right)^{\!\!-1},
	\label{beta-def}
	\ee
and (\ref{bgs=1}) gives
	\[\bg_s=\fs(|a_z|k)\left[\big\{\beta(\fz_x)-1\big\}j_x\,\bfe_x+
	\big\{\beta(\fz_y)-1\big\}j_y\,\bfe_y\right].\]	
In particular, if $\fz_x=\fz_y$, so that the point scatterer is uniaxial or isotropic, we have 
	\be
	\bg_s=\fs(|a_z|k) [\beta(\fz_x) -1]\,\vec j,
	\label{gs-iso=}
	\ee 
where $\vec j:=j_x\bfe_x+j_y\bfe_y$. 
	\end{enumerate}
	
\begin{figure}
        \begin{center}
        \includegraphics[scale=.55]{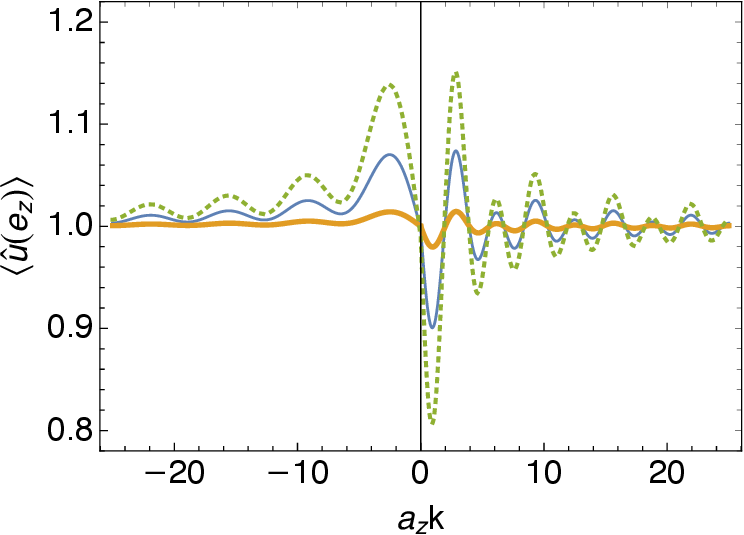}~~~~~~~\includegraphics[scale=.55]{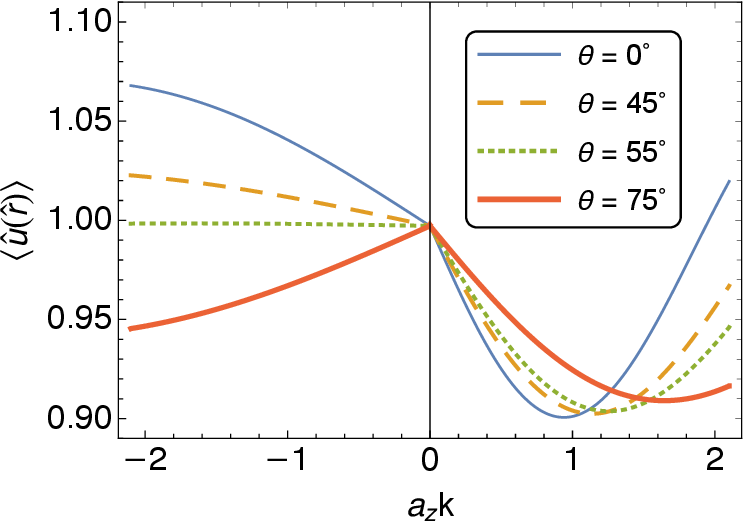}
        \caption{Plots of the normalized time-averaged intensity as a function of $a_zk$ for $\fz_x=\fz_y$. Left panel gives the plots of $\br \hat u(\bfe_z)\kt$ for $\fz_x k^3=0.2$ (thick orange curve), 1.0 (thin blue curve), and $2.0$ (dotted green curve). The right panel shows the plots of $\br \hat u(\hat\bfr)\kt$ for $j_z=0$, $\fz_xk^3=1$, and $\vartheta=0^\circ, 45^\circ, 55^\circ$, and $75^\circ$. For $|a_zk|\lesssim 0.9$ and $0\leq\vartheta\lesssim 55^\circ$, the value of $\br \hat u(\hat\bfr)\kt$ determines $a_zk$ uniquely.}
        \label{fig2}
        \end{center}
        \end{figure}

Let $\br u(\bfr)\kt$ and $\br u_0(\bfr)\kt$ denote the time-averaged energy density of the detected wave at $\bfr$ in the presence and absence of point scatterers, respectively. According to (\ref{cE-detected-s-dipole}), the ratio of these quantities, which equals the ratio of the time-averaged intensities of these waves \cite{griffiths-EM}, is given by
	\be
	\frac{\br u(\bfr)\kt}{\br u_0(\bfr)\kt}=\br \hat u(\hat\bfr)\kt:=
	\frac{\big|\hat\bfr\times(\hat\bfr\times(e^{-ik\bfa\cdot\hat r}\bj+\bg_s)
	\big|^2}{\big|\hat\bfr\times(\hat\bfr\times \bj)\big|^2}.
	\label{u-def}
	\ee
Substituting (\ref{fs-def}), (\ref{beta-def}), and (\ref{gs-iso=}) in this equation, we find $\br \hat u(\hat\bfr)\kt$ for $\fz_x=\fz_y$. This in particular implies 
	\be
	\br \hat u(\hat\bfr)\kt=\left\{\begin{array}{ccc}
	\big|e^{ika_z}\fs(|a_z|k) [\beta(\fz_x) -1]+1\big|^2 &\for & \hat\bfr=\bfe_z,\\[3pt]
	\big|e^{ika_z\cos\vartheta}\fs(|a_z|k) [\beta(\fz_x) -1]+1\big|^2 &\for & j_z=0.
	\end{array}\right.
	\label{intensity}
	\ee
Fig.~\ref{fig2} shows the graphs of the normalized time-averaged intensity (\ref{intensity}) measured by detectors located at $z=+\infty$ as a function of $a_zk$. As expected, $\br \hat u(\bfe_z)\kt$ tends to 1 as $|a_zk|$ grows, and its deviation from 1 is more pronounced for larger values of $|\fz_x| k^3$.  For $\fz_xk^3=1$, $|a_zk|\lesssim 0.9$, and $0\leq\vartheta\lesssim 55^\circ$, $\br\hat u(\hat\bfr)\kt$ turns out to be a one-to-one function of $a_zk$. This shows that one can in principle use the value of the normalized time-averaged intensity for sufficiently low-energy waves to determine the relative position of the source with respect to the point scatterer or vice versa, if one can identify the line joining them (i.e., the $z$ axis).\footnote{Plotting the graph of $\br\hat u(\hat\bfr)\kt$ for different real and complex values of $\fz_xk^3$, we have checked that this feature is not sensitive to the value of $\fz_xk^3$.} This simple example suggests the possibility of using the exact solution of the radiation problem in the presence of point scatterers for the purpose of addressing the inverse problem of locating the scatterers using the data on their response to the incident radiation, which is a problem of great practical importance.

\section{Conclusion}
\label{S6}
Multi-dimensional generalizations of the transfer matrix of scattering theory in one dimension has been developed and utilized since the 1980's basically for the purpose of numerical investigation of wave propagation in stratified media \cite{pendry-1984}. The basic idea behind these developments is to dissect the medium into a large number of thin layers along a propagation axis, discretize the transverse degrees of freedom in each layer, assign a transfer matrix to each layer (which is a matrix relating the amplitude of the wave at the points representing one of the two large boundaries of the layer to the other), and multiply them according to a particular composition rule to obtain the transfer matrix for the bulk. Recently, we have introduced a fundamental notion of transfer matrix for scalar \cite{pra-2021} and electromagnetic \cite{pra-2023} waves whose definition does not require the slicing or discretization of the medium. This notion forms the basis of a dynamical formulation of the stationary scattering that allows for analytic calculations and is particularly effective in dealing with point interactions. 

In the present article, we explore the utility of the fundamental transfer matrix in the study of the problem of radiation in a general linear scattering medium. This leads to a general method of solving this problem. Using this method we have shown that the electric field of the wave emitted by an oscillating source has the form
	\[\boldsymbol{\mathsf{E}}(\bfr,t)=\frac{k\,e^{i(kr-\omega t)}}{4\pi i\sqrt{\varepsilon_0}\,r}
	\:\,\hat\bfr\times(\hat\bfr\times\bg)~~~\for~~~r\to\infty,\]
where $\bg$ is a vector belonging to the $x$-$y$ plane that stores all the information about the current density characterizing the source and the permittivity and permeability tensors of the medium. We have provided the following procedure for the calculation of $\bg$.
	\begin{enumerate}
	\item Determine the evolution operator $\widehat\bcU(+\infty,z)$ given by (\ref{dyson}) which specifies the dynamics of the non-unitary effective quantum system corresponding to the interaction-picture Hamiltonian $\widehat\bcH(z)$.
	\item Calculate the fundamental transfer matrix $\widehat\bM$ and the $4$-component field $\bD$ which are respectively given by (\ref{M-def}) and (\ref{D=}).
	\item Solve the integral equation (\ref{C-minus=1}) for the $4$-component field $\bC_-$.
	\item Read off the expressions for $\vec C_-^-$, $\vec G_-$, and $\bg_-$ using (\ref{2-comp-Cpm}), (\ref{vG-pm-def}), and (\ref{gpm-def=0}), and identify $\bg$ with $\bg_-$.	
	\end{enumerate}

We have successfully applied our method to describe the radiation of an oscillating source placed in a medium consisting of a regular or irregular planar array of nonmagnetic, possibly anisotropic and active or lossy point scatterers. For this system, which is relevant to the study of nanoparticles having extremely large refractive indices \cite{cao}, the determination of $\bg$ requires the solution of a linear system of $2N$ algebraic equations (\ref{system}), where $N$ is the number of point scatterers, and the evaluation of the integral in (\ref{h-def}) which seems not to admit an explicit expression in terms of the known functions. This is clearly not a major problem, for we can compute it numerically. We can also find the numerical solution of  (\ref{system}) for large arrays consisting as many as hundreds of point scatterers. 

A distinctive feature of our treatment of point scatterers is that it avoids the singularities of their traditional treatments \cite{VCL,calla-2014}. This is among the main difficulties in dealing with the radiation problem in the presence of point scatterers which we have been able to circumvent. 


\vspace{12pt}

\noindent {\bf Acknowledgements}:
This work has been supported by the Scientific and Technological Research Council of T\"urkiye (T\"UB\.{I}TAK) in the framework of the project 120F061 and by Turkish Academy of Sciences (T\"UBA).

\ed